\documentclass[iop,apj,twocolappendix]{emulateapj}
\usepackage{amsmath,amssymb,amstext}

\usepackage[breaklinks,colorlinks,citecolor=blue,linkcolor=magenta]{hyperref} 

\usepackage[all]{hypcap} 
\usepackage{multirow}
\usepackage{booktabs}
\usepackage{enumitem}
\usepackage{rotating}

\usepackage{aas_macros}
\usepackage{natbib}
\shorttitle{Stellar Populations and Star Formation History of DDO~68}

\shortauthors{Sacchi et al.}

\begin{document}

\title{Stellar populations and Star Formation History of the Metal-Poor Dwarf Galaxy DDO~68 \footnotemark[$\star$]} \footnotetext[$\star$]{Based on observations obtained with the NASA/ESA Hubble Space Telescope at the Space Telescope Science Institute, which is operated by the Association of Universities for Research in Astronomy under NASA Contract NAS5-26555.}%
\author{E. Sacchi$^{1,2}$, F. Annibali$^{1,2}$, M. Cignoni$^{3,4,5}$, A. Aloisi$^{3}$, S. T. Sohn$^{6}$, M. Tosi$^{2}$, R. P. van der Marel$^{3}$,\\A. J. Grocholski$^{7}$ and B. James$^{8}$}
\affil{$^{1}$Dipartimento di Fisica e Astronomia, Universit\`a degli Studi di Bologna, Viale Berti Pichat 6/2, I-40127 Bologna, Italy\\
$^{2}$INAF-Osservatorio Astronomico di Bologna, Via Ranzani 1, I-40127 Bologna, Italy; elena.sacchi5@unibo.it\\
$^{3}$Space Telescope Science Institute, 3700 San Martin Drive, Baltimore, MD 21218, USA\\
$^{4}$Dipartimento di Fisica, Universit\`a di Pisa, Largo Bruno Pontecorvo, 3, 56127 Pisa, Italy;\\
$^{5}$INFN, Sezione di Pisa, Largo Pontecorvo 3, 56127 Pisa, Italy\\
$^{6}$Department of Physics and Astronomy, The Johns Hopkins University, Baltimore, MD 21218, USA\\
$^{7}$Department of Physics \& Astronomy, Louisiana State University, Baton Rouge, LA 70803, USA\\
$^{8}$Institute of Astronomy, University of Cambridge, Madingley Road, Cambridge, CB3 0HA}

\begin{abstract}
We present the star formation history of the extremely metal-poor dwarf galaxy DDO~68, based on our photometry with the Advanced Camera for Surveys. With a metallicity of only $12+\log(O/H)=7.15$ and a very isolated location, DDO~68 is one of the most metal-poor galaxies known. It has been argued that DDO~68 is a young system that started forming stars only $\sim 0.15$~Gyr ago. Our data provide a deep and uncontaminated optical color-magnitude diagram that allows us to disprove this hypothesis, since we find a population of at least $\sim 1$~Gyr old stars. The star formation activity has been fairly continuous over all the look-back time. The current rate is quite low, and the highest activity occurred between 10 and 100 Myr ago. The average star formation rate over the whole Hubble time is $\simeq 0.01$~M$_{\odot}$~yr$^{-1}$, corresponding to a total astrated mass of $\simeq 1.3 \times 10^8$~M$_{\odot}$. Our photometry allows us to infer the distance from the tip of the red giant branch, $D = 12.08 \pm 0.67$~Mpc; however, to let our synthetic color-magnitude diagram reproduce the observed ones we need a slightly higher distance, $D=12.65$~Mpc, or $(m-M)_0 = 30.51$, still inside the errors of the previous determination, and we adopt the latter. DDO~68 shows a very interesting and complex history, with its quite disturbed shape and a long Tail probably due to tidal interactions. The star formation history of the Tail differs from that of the main body mainly for an enhanced activity at recent epochs, likely triggered by the interaction.
\end{abstract}

\keywords{galaxies: dwarf -- galaxies: evolution -- galaxies: individual (DDO~68) -- galaxies: irregular -- galaxies: starburst -- galaxies: stellar content}

\maketitle

\section{Introduction}
\setcounter{footnote}{7}
According to hierarchical galaxy formation models, dwarf ($M_{\star}\lesssim 10^9$~M$_{\odot}$) galaxies are the first systems to form, providing the building blocks for the formation of more massive systems through continuous merging and accretion. As a consequence, present-day dwarfs may have been sites of the earliest star formation (SF) activity in the Universe. On the other hand, in the past, given their high gas content and very blue integrated colors, indicative of the prevalence of young stellar populations, the most metal-poor $(12+\log(O/H)\lesssim 7.6$, corresponding to $Z\lesssim 1/15$~Z$_{\odot}$\footnote{Adopting A(O)=$8.76\pm 0.07$ from \cite{Caffau2008}}) dwarf irregular (dIrr) and blue compact dwarf (BCD) galaxies have been often suggested to be ``primeval'' galaxies, experiencing their first burst of star formation, with ages $\lesssim 100-500$~Myr \citep{Izotov1999,Pustilnik2005,Pustilnik2008}. 
Only by performing resolved stellar population studies we can properly investigate whether dIrrs/BCDs do indeed host only young stars. All dIrr/BCD galaxies resolved and studied so far with the Hubble Space Telescope (\textit{HST}) have been found to harbor stars as old as the look-back time sampled by the depth of the photometry, i.e. $\sim 1$~Gyr and older \citep{Tolstoy1998,Schulte-Ladbeck2002,Izotov2002,Tosi2009,Tolstoy2009}. The significance of such studies is clearly illustrated in the long-standing controversial case of I~Zw~18, the most metal-poor prototype of the BCD class with $12 + \log(O/H) = 7.2$ \citep{Skillman1993}. Imaging with the Advanced Camera for Surveys (ACS) on board of \textit{HST} performed by \cite{Aloisi2007} provided a deep and uncontaminated color-magnitude diagram (CMD) that strongly indicated the presence of a previously undetected red giant branch (RGB), thus ruling out its former classification as a truly primordial galaxy \citep{Izotov2004}.

Over the past years, several new dIrr/BCDs with extremely low metallicities and physical properties similar to I~Zw~18 have been discovered and controversially regarded as ``genuine'' young galaxies in the nearby Universe due to a lack of detailed information on their resolved stellar population ages. One of the most recent cases is Leo P, discovered by \cite{Giovanelli2013} within the ALFALFA survey, with a metallicity from H~II region spectra of $12+\log(O/H) = 7.17 \pm 0.04$ \citep{Skillman2013}. \cite{McQuinn2015} observed it with \textit{HST} and indisputably found RR Lyrae stars, i.e. stars at least 10 Gyr old, from which they also inferred a robust distance estimate of $1.62 \pm 0.15$~Mpc.

In this paper we present the interesting case of DDO~68 (UGC~5340), which holds the same record-low metallicity as I~Zw~18 and Leo~P, $12 + \log(O/H) = 7.15 \pm 0.04$, from longslit-spectroscopy of the ionized gas in its H~II regions \citep{Izotov2009}. 
After deriving a distance of $12.74\pm 0.27$~Mpc, \cite{Cannon2014} estimated for DDO~68 a dynamical mass of $\mathrm{M_{dyn}} \sim 5.2\times 10^9$~M$_{\odot}$ within 11 kpc, and an H~I mass of $\mathrm{M_{H~I}} = (1.0 \pm 0.15) \times 10^9$~M$_{\odot}$. Their  distance, although larger than the $\sim 9.9$~Mpc predicted by its recession velocity ($v = 502 \pm 4$~km/s), 6.5~Mpc from the apparent magnitudes of the brightest stars \citep{Pustilnik2005}, and 12.0~Mpc derived from the Tip of the Red Giant Branch (TRGB) \citep{Tikhonov2014}, places DDO~68 a factor 1.5 closer than I~Zw~18, close enough to resolve its old stars with \textit{HST}.

DDO~68 is especially appealing as candidate young galaxy because it is located at the periphery of the nearby Lynx-Cancer void \citep{Pustilnik2011} and is at least 2 Mpc away from the closest bright galaxies ($L > L_{\star}$, where $L_{\star}$ corresponds to $M_B=-19.6$ for $H=72$~km~s$^{-1}$~Mpc$^{-1}$). Within the framework of hierarchical formation scenarios, the evolution of galaxies formed in voids could be quite different than in higher density environments because of a much lower probability of being subject to those processes (i.e. encounters and mergers) that are important in shaping and accelerating the evolution of galaxies \citep{Peebles2001,Gottlober2003,Rojas2004,Rojas2005,Hoeft2006}. As a consequence, some void galaxies may have survived in their nascent state of gaseous protogalaxies until the present epoch, and may have just started forming stars as a consequence of a recent disturbance.

\begin{figure*}
\centering
\includegraphics[trim=0 0 0 -0.5cm, clip, width=\linewidth]{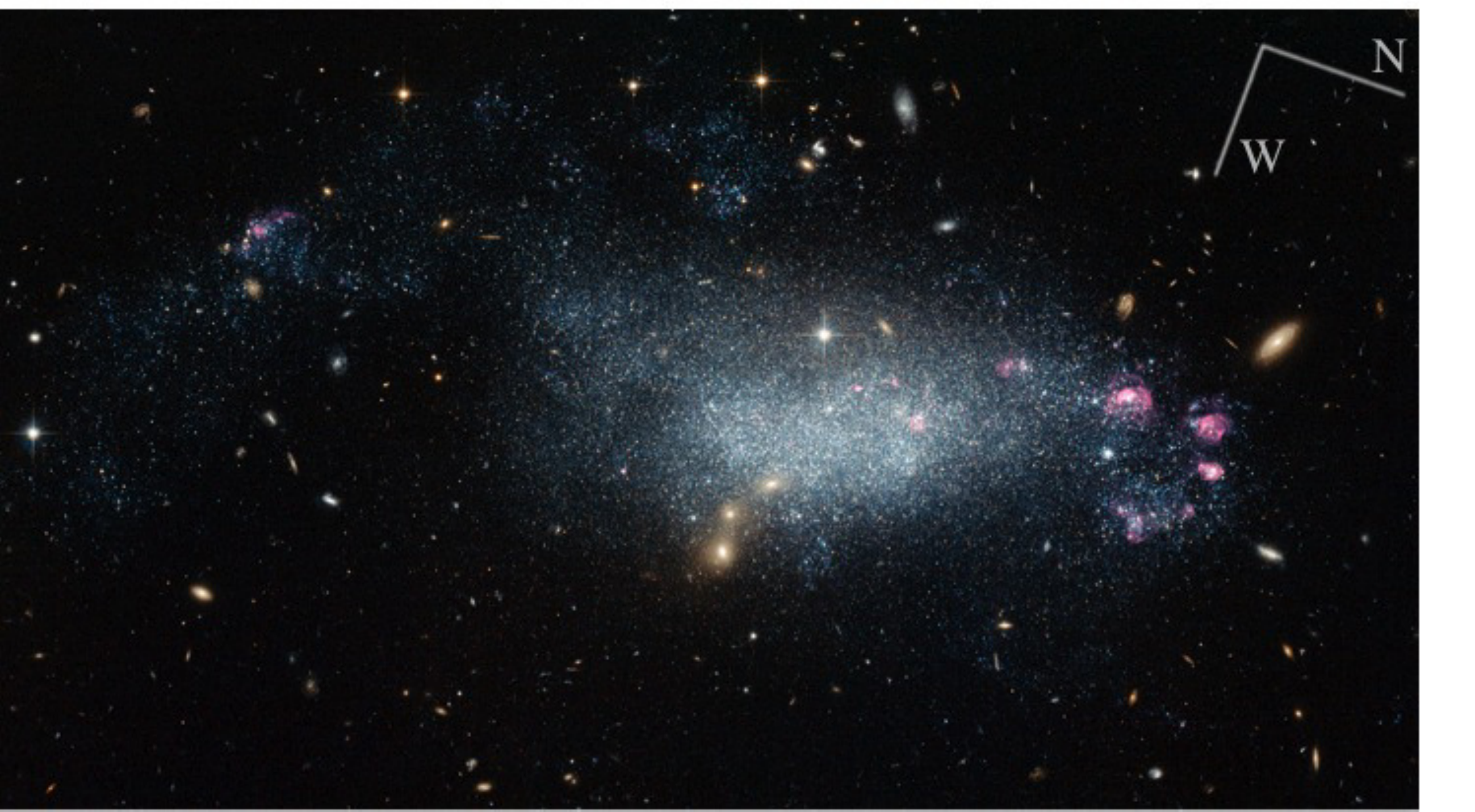}
\caption{ACS/WFC color-combined image of DDO~68 (blue=broad $V$, green=$I$, red=$H\alpha$)}
\label{3col}
\end{figure*}

\begin{figure*}
\centering
\includegraphics[trim=0 -0.5cm -0.75cm -0.5cm, clip, width=\linewidth]{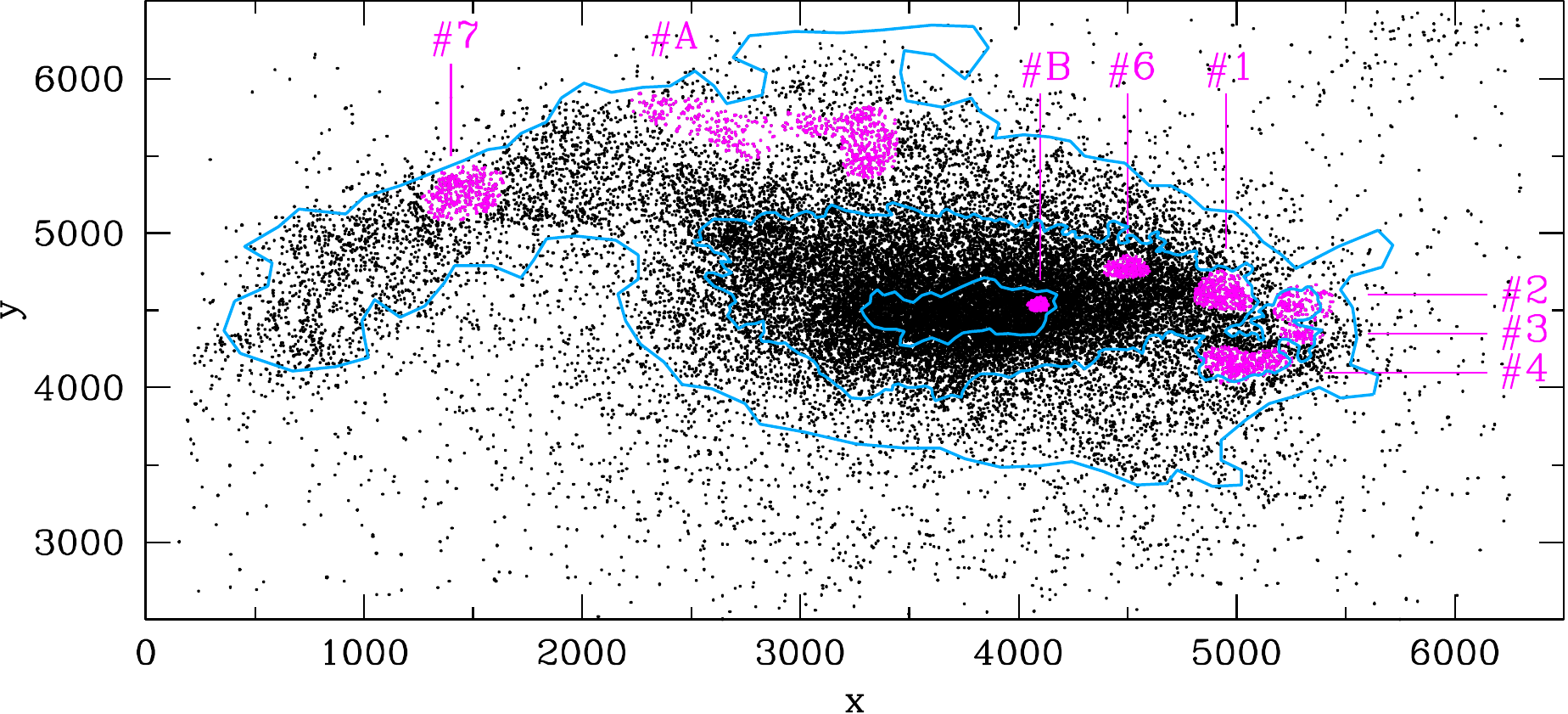}
\caption{Map of the stars with the isophotal contours superimposed (light blue lines) which were used to identify the 4 regions for the analysis of the galaxy SFH. Region 1 is the most central one, and Regions 2, 3 and 4 are increasingly more external. The magenta points are the stars within the H II regions labeled as \cite{Pustilnik2005} and \cite{Izotov2009} except for A and B that they don't take into account in their analysis.}
\label{map}
\end{figure*}

DDO~68 presents a very irregular optical morphology, with a long curved Tail on the South and a ring-like structure at the Northern edge consisting of 5 separated H II regions (see Figs. \ref{3col} and \ref{map}). The neutral gas morphology and kinematics as inferred from H~I maps by \cite{Stil2002a,Stil2002b} are very disturbed as well. DDO~68 is also an outlier on the mass-metallicity relationship \citep{Pustilnik2005,Berg2012}, being overly massive compared to other systems with comparable metallicity. All these characteristics suggest that DDO~68 may have undergone an interaction event. However, the closest known neighbor galaxy, the dwarf UGC~5427, is at a projected distance of $\sim 200$~kpc. There is no evidence for a physical association between the two galaxies, and with $M_B = −14.5$, UGC~5427 has a too low mass to have significantly affected DDO~68. So, the absence of an easily identifiable companion to DDO~68 leaves the question of how its recent SF was triggered open. From the analysis of the spatial distribution of stars resolved in \textit{HST}/ACS imaging, \citep{Tikhonov2014} argued that DDO~68 consists in fact of two different systems: a central most massive body and an elongated arc-shaped body (DDO~68~B). In a recent work based on Very Large Array (VLA) observations, \cite{Cannon2014} claimed to have found an object with the same systemic velocity of DDO~68. This gas-rich galaxy (DDO~68 C), has a total neutral hydrogen mass of $\mathrm{M_{H~I}} = (2.8 \pm 0.5) \times 10^7$~M$_{\odot}$ and lies at a distance of $\sim 42$~kpc from DDO~68. The detection of a bridge of low surface brightness gas connecting DDO~68 and DDO~68 C may suggest an interaction between the two objects.

In this paper we present the star formation history (SFH) of DDO~68 based on deep \textit{HST}/ACS data. The derived CMD, which reaches down to $\sim 1$~mag below the TRGB, implies that DDO~68 started its star formation at least $\sim 1-2$~Gyr ago (and possibly up to a whole Hubble time) disproving the hypothesis that it is a truly primeval galaxy of recent formation.
We analyzed the populations of stars resolved with HST in and around DDO~68 and applied the synthetic CMD method \citep{Tosi1991} to infer its SFH. We also provide a new distance estimate, based both on the TRGB and on the whole CMD fitting.\\

\section{Observations and Data Reduction}
\label{obs}
\begin{figure}
\includegraphics[width=\columnwidth]{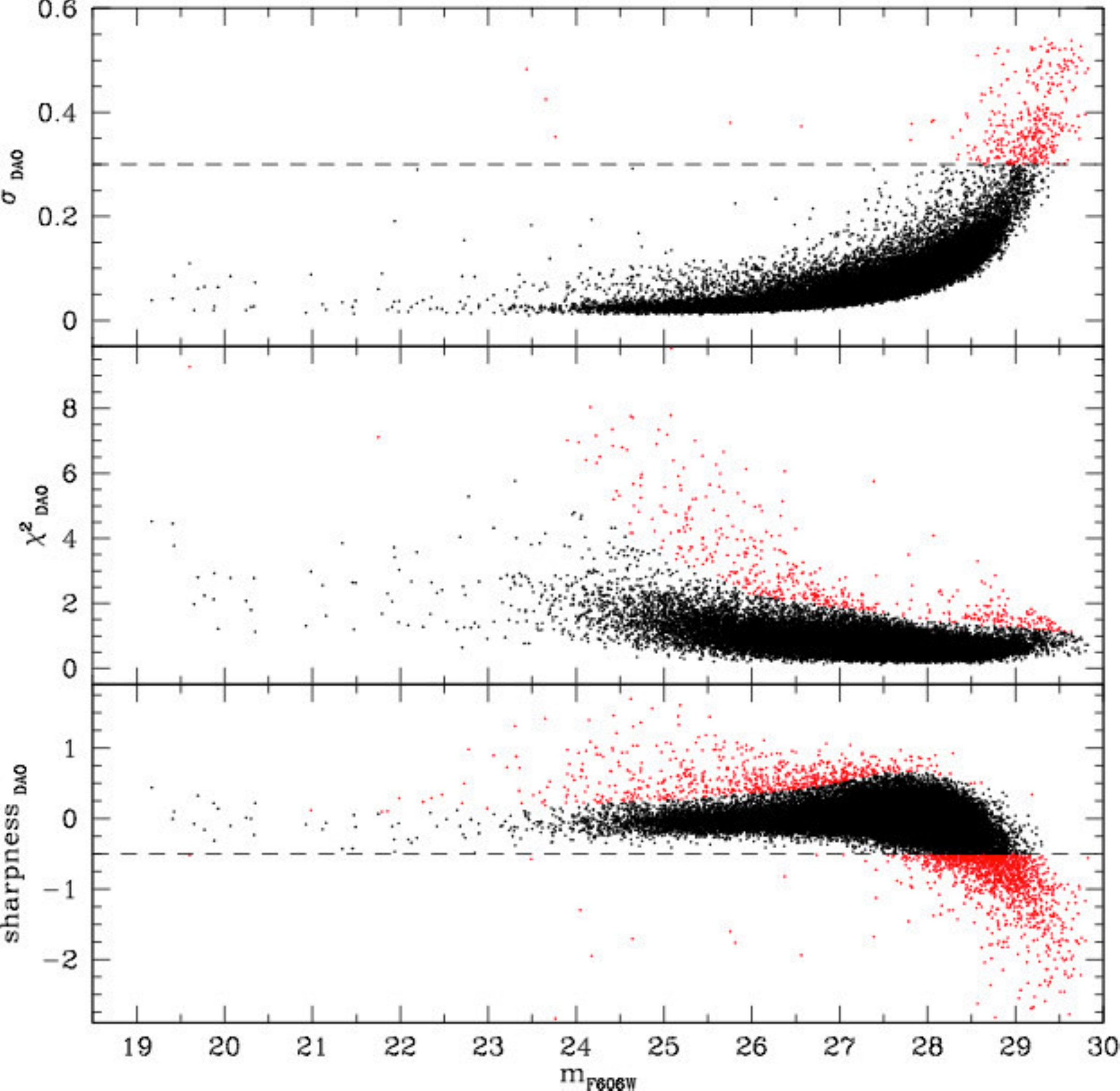}
\includegraphics[width=0.99\columnwidth]{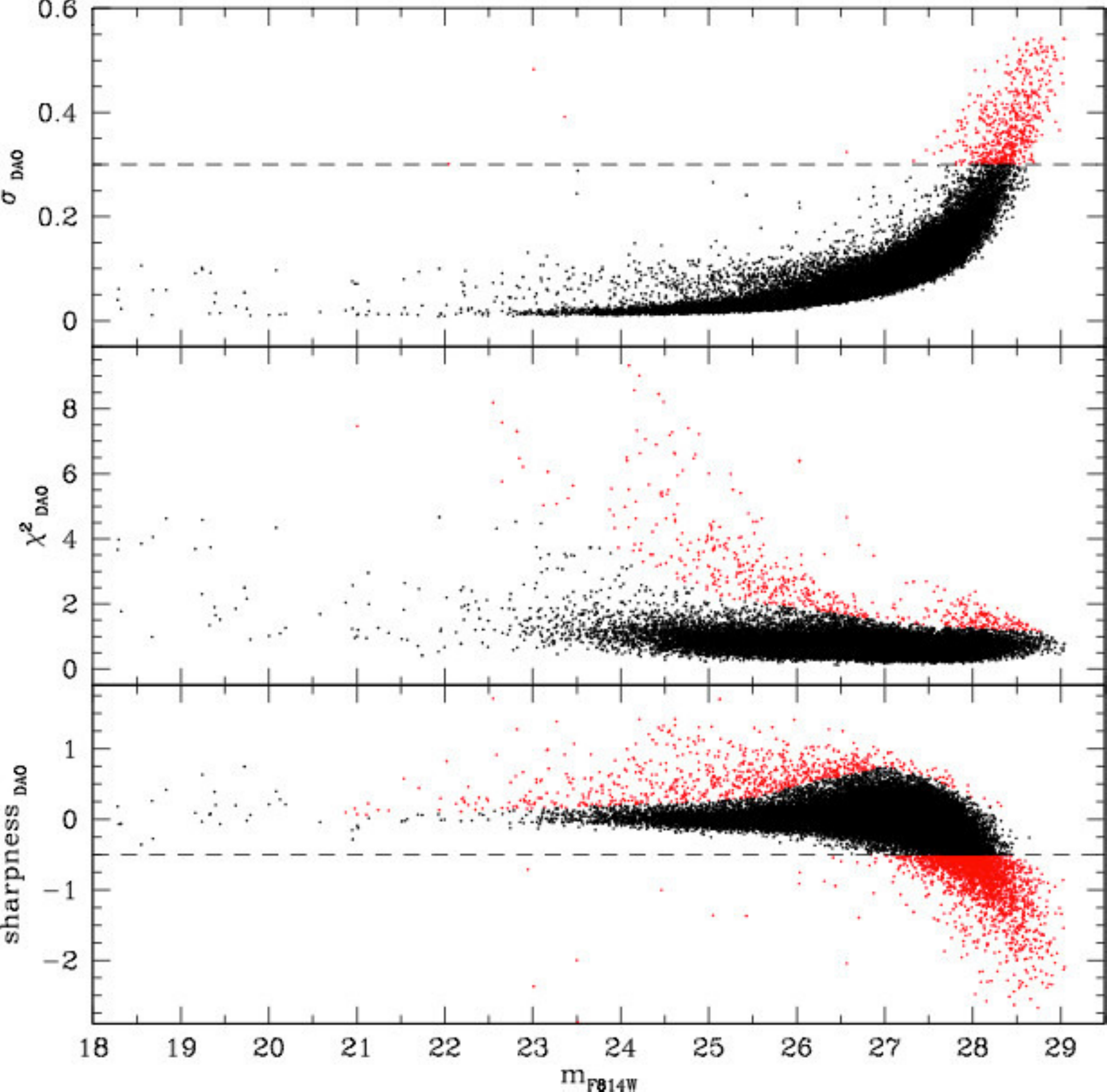}
\caption{Selection cuts applied to the photometric error $\sigma_{\mathrm{DAO}}$ (top panels), $\chi^2$ (middle panels) and sharpness (bottom panels) for the two filters. The red and black points are, respectively, the discarded and the retained objects.\\}
\label{cuts}
\end{figure}
Deep imaging with the ACS Wide Field Channel (WFC) of DDO~68 was performed on 2010 April 27 and May 2 (GO program 11578; PI Aloisi) using the F606W ($\sim$ broad $V$) and F814W ($\sim I$) broad-band filters and the F658N (H$\alpha$) narrow-band filter. The total integration time is $\sim 2400$~s in F658N, and $\sim 6940$~s in both F606W and F814W. In addition, coordinated observations were made in parallel with the UV-optical channel (UVIS) of the Wide Field Camera 3 (WFC3), in both F606W and F814W filters. The results from the latter are not discussed here and will be presented in a subsequent publication.

We have observed DDO~68 with standard dither pattern in both filters to fill the gap between the two CCDs of the ACS and improve the sampling of the point-spread function (PSF). All images were processed with the latest version of CALACS, the ACS pipeline. We applied to them all the routine developed by \cite{Anderson2010} to corrected for Charge Transfer Efficiency (CTE). Eventually, we combined all the images in the same filter into a single frame using the MULTIDRIZZLE software package \citep{Fruchter2009}. MULTIDRIZZLE allows to fine-tune the alignment of the images, correct for small shifts, rotations and geometric distortions between the images, and remove cosmic rays and bad pixels. The final $V$ and $I$ images were eventually resampled to a pixel size of 0.033 arcsec (0.68 times the original ACS-WFC pixel size). Figure \ref{3col} shows a three-color combined image of DDO~68.

We have performed PSF-fitting photometry on the $V-$ and $I-$ band images using the stand-alone versions of DAOPHOT and ALLSTAR \citep{Stetson1987}. To create the PSF models, we selected $\sim 174$ and $\sim 265$ bright, relatively isolated stars in $V$ and $I$, respectively, with good spatial coverage across the entire image. The PSF was modeled with analytic MOFFAT (F606W) and PENNY (F814W) functions plus second-order additive corrections as a function of position derived from the residuals of the fit to the PSF stars. Next, we ran ALLSTAR to fit the PSF models to the sources detected independently in the $V$ and $I$ images.
To push the photometry as deep as possible, we repeated the procedure on the subtracted images, allowing for the detection of faint companions near brighter stars. The $V$ and $I$ catalogs were finally cross-correlated adopting a matching radius of 1 pixel, providing a final matched catalog of $\sim 35,000$ objects.

The calibration of the instrumental magnitudes into the \textit{HST} VEGAMAG system was performed by applying the \cite{Bohlin2012} time-dependent ACS zero points provided by STScI\footnote{\url{http://www.stsci.edu/hst/acs/analysis/zeropoints}}. To determine the aperture corrections to the conventional 0\farcs5 radius aperture, we performed photometry on several bright and isolated stars in our images\footnote{The correction from 0\farcs5 to an infinite aperture was done using the old \cite{Sirianni2005} values instead. However, according to \cite{Bohlin2012} the new Encircled Energy values are within 1-2\% of the corresponding \cite{Sirianni2005} results, therefore we expect a negligible effect on our photometry, considering also the typical size of the photometric errors.}.

In order to clean the photometric catalog from spurious objects, we applied selection cuts on the DAOPHOT parameters $\sigma$ (photometric error), $\chi^2$ and sharpness, as shown in Fig. \ref{cuts}: to retain a star, we required that $\sigma_V < 0.3$, $\sigma_I < 0.3$, sharp$_V >-0.5$, sharp$_I >-0.5$. Furthermore, we applied an additional cut in sharpness and $\chi^2$ following the upper envelopes of the distributions (see Fig. \ref{cuts}).

A visual inspection in the images of the rejected objects, coupled with an analysis of their radial profiles, revealed that the majority of them are background galaxies, unresolved stellar clusters or blends of two or more stars. The same inspection was performed for objects brighter than $\sim 23$~mag in $V$ or $I$, independently of their $\chi^2$ and sharpness. After these selections, we were left with $\sim 30,000$ objects.\\

\section{Incompleteness and Errors}
Incompleteness and errors of our photometry have been evaluated performing artificial star tests on the images. To this aim, we added artificial stars created with the adopted PSF to the images and exploited the DAOPHOT/ADDSTAR routine, that simulates real stars adding the appropriate Poisson noise to the artificial stars, and allows to cover the desired range of positions and magnitudes. The range of magnitudes covered by the artificial stars was chosen to match the data. We divided the images into grids of $30\times 30$~pixel$^2$ cells and placed one artificial star in each cell at each run, to efficiently compute the completeness without altering the crowding on the images. To avoid overcrowding, we also imposed that no two artificial stars laid within 20 pixels of each other, since the PSF radius is 15 pixels. The images were fully sampled because we let the starting position of the grid and the position of the artificial star within each cell vary appropriately.  We performed PSF fitting photometry of the images with the added artificial stars with the same procedure as for the real data described in Section \ref{obs}. We then cross-correlated stars in this output catalog with those in the combined catalog that includes both the original photometric catalog and the input artificial star catalog. An input artificial star was considered ``lost'' when it was not recovered in the output catalog, or when its recovered magnitude differed from the input value by more than 0.75 mag. We then cleaned the artificial star catalog applying the same criteria in photometric error, $\chi^2$ and sharpness used in the real data. In total, we simulated $\sim 1,800,000$ stars.\\
\begin{figure*}
\includegraphics[width=\columnwidth]{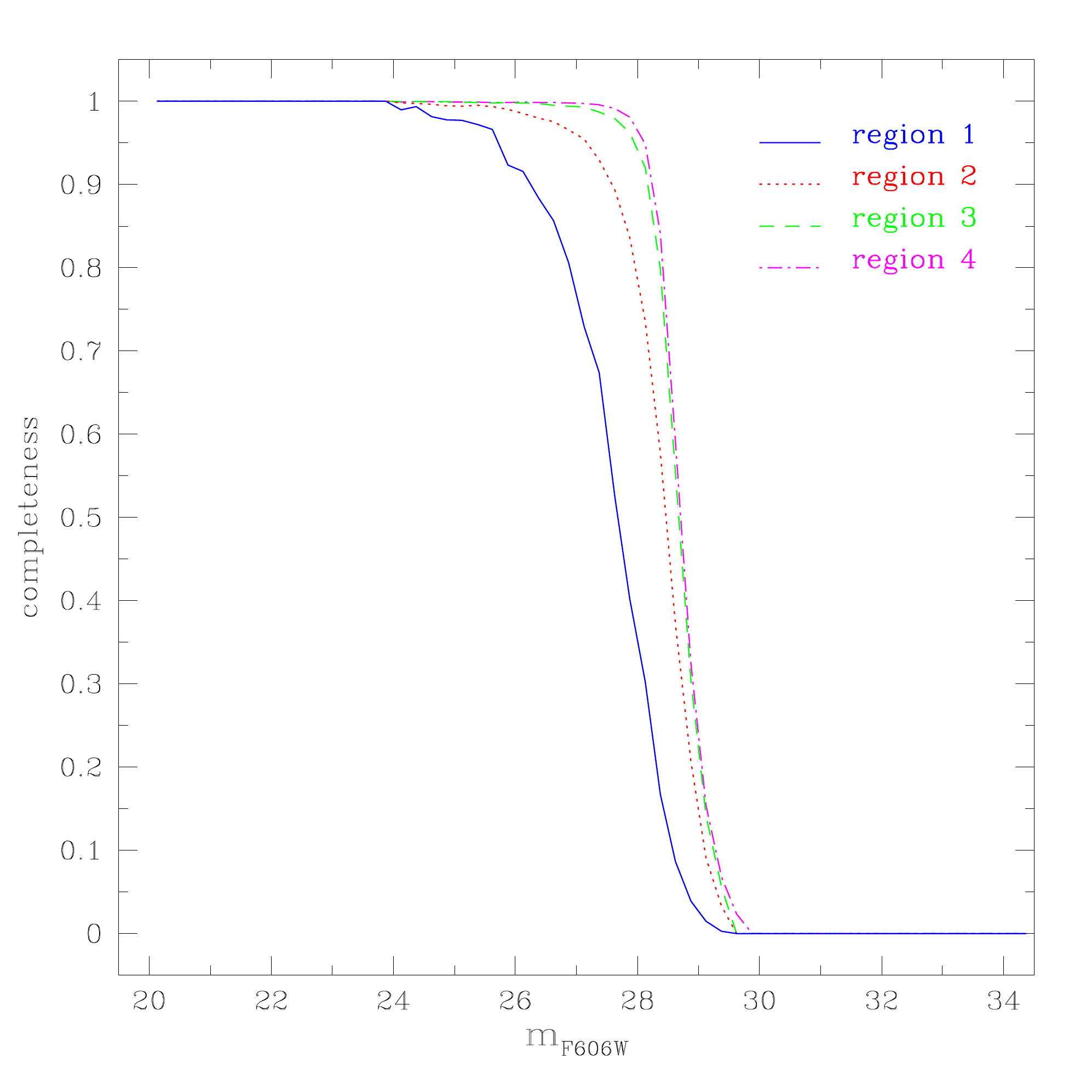}
\includegraphics[width=\columnwidth]{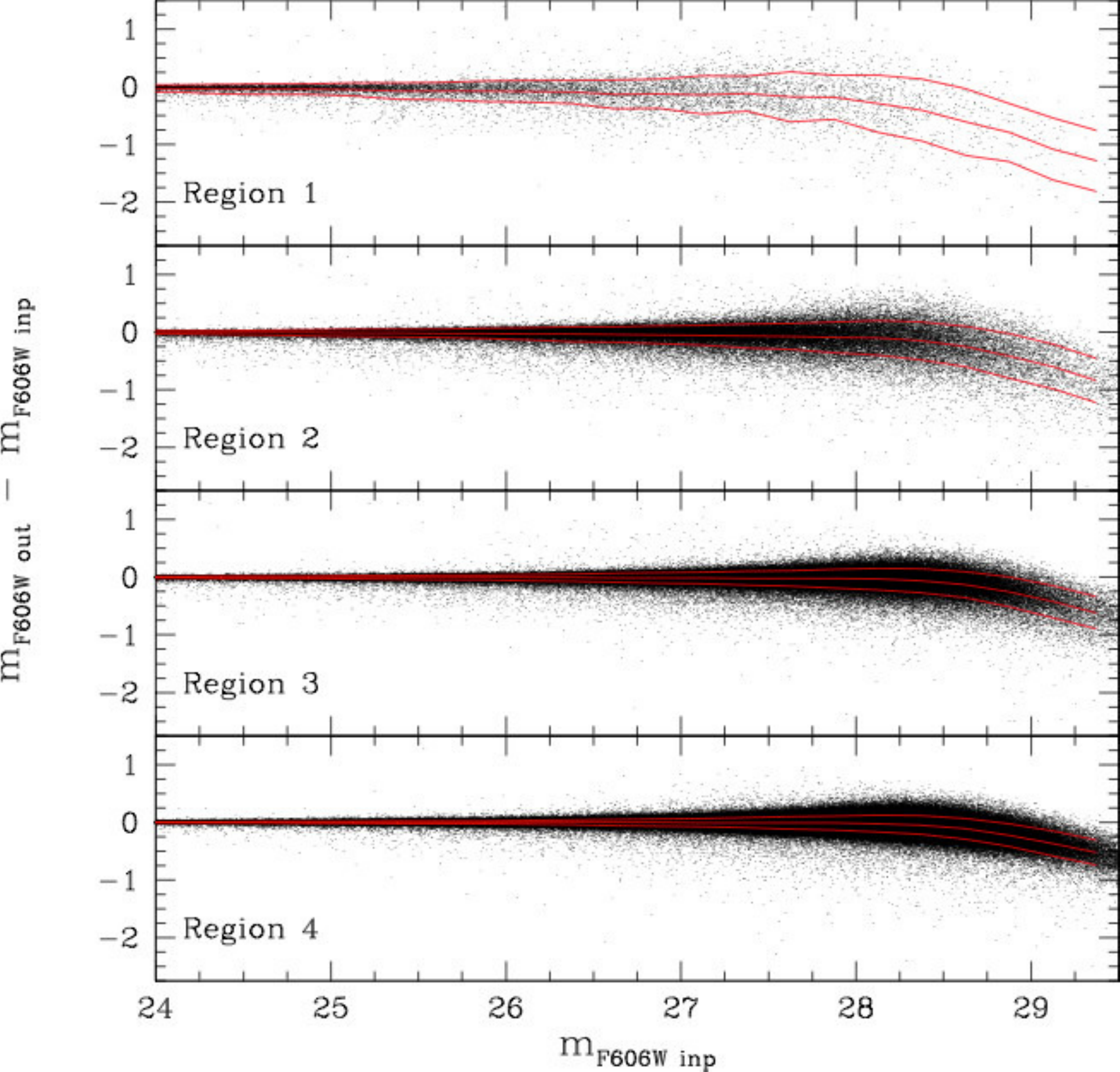}
\includegraphics[width=\columnwidth]{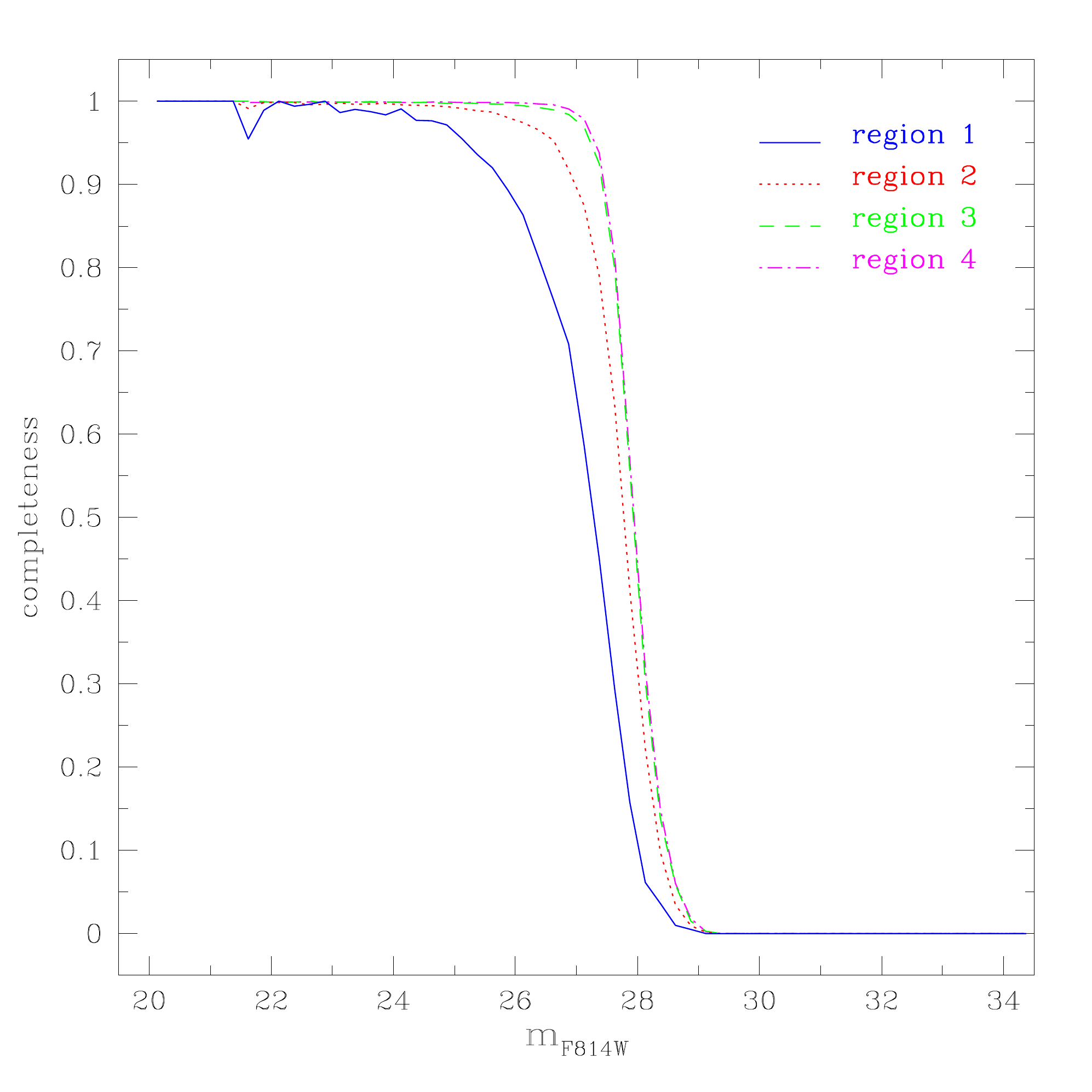}
\hfill
\includegraphics[width=\columnwidth]{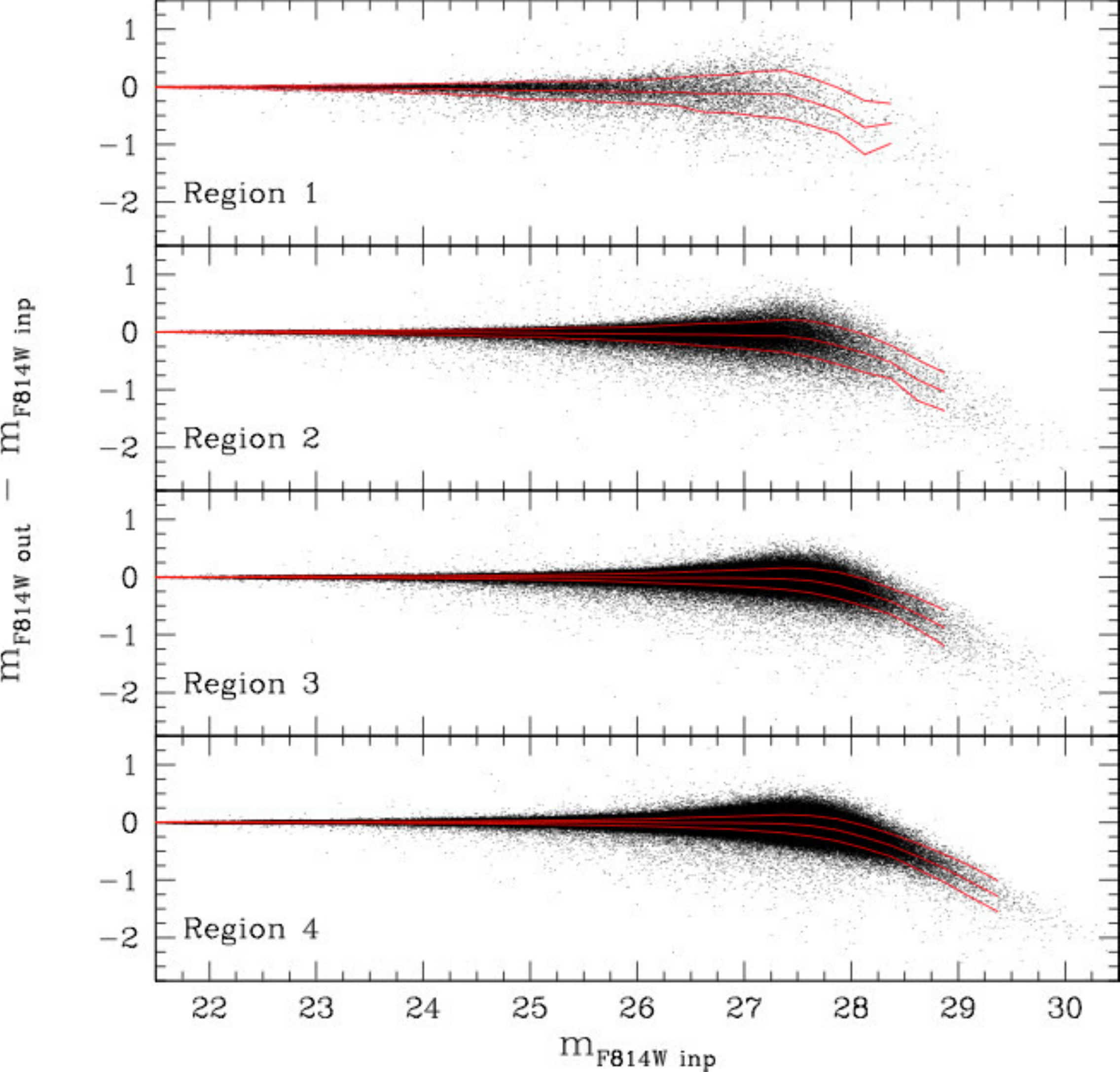}
\caption{Completeness (left panels) and photometric errors (right panels) in $V$ and $I$ from the artificial star tests. The solid red curves indicate the mean of the $\Delta$mag distribution (central line) and the $\pm 1\sigma$ standard deviations.}
\label{complerr}
\end{figure*}
The completeness of our photometry was taken as the ratio of the number of recovered artificial stars over their input number in each magnitude bin. In order to account for variable crowding within the galaxy, as well as for gradients in the stellar populations (see Sections \ref{sec_cmd} and \ref{pop_distr}), we identified 4 regions within DDO~68 using isophotal contours, with Region 1 being the most central one and Region 4 the most external one. The contours that identify the four regions are overplotted on Fig. \ref{map} showing the spatial map of the resolved stars.

In Fig. \ref{complerr}, left panels, we show the behavior of the $V$ and $I$ completeness as a function of magnitude for the four regions. As expected, Region 1 is the most incomplete one due to the largest crowding, with the completeness starting to drop at $V \sim 26$ and $I \sim 25$. On the other hand, the photometry is $\sim$ 100\% complete down to $V \sim 28$ and $I \sim 27$ in the most external Regions 3 and 4. The behavior of the output-input magnitudes, displayed in the right panels of Fig. \ref{complerr}, provides a more realistic estimate of the photometric errors than $\sigma_{\mathrm{DAO}}$. The increasing deviation of the average $\mathrm{m_{out}-m_{inp}}$ towards negative values indicates the increasing effect of blending toward fainter magnitudes.\\

\section{The Color-Magnitude Diagram} \label{sec_cmd}
\begin{figure*}
\centering
\includegraphics[width=0.99\linewidth]{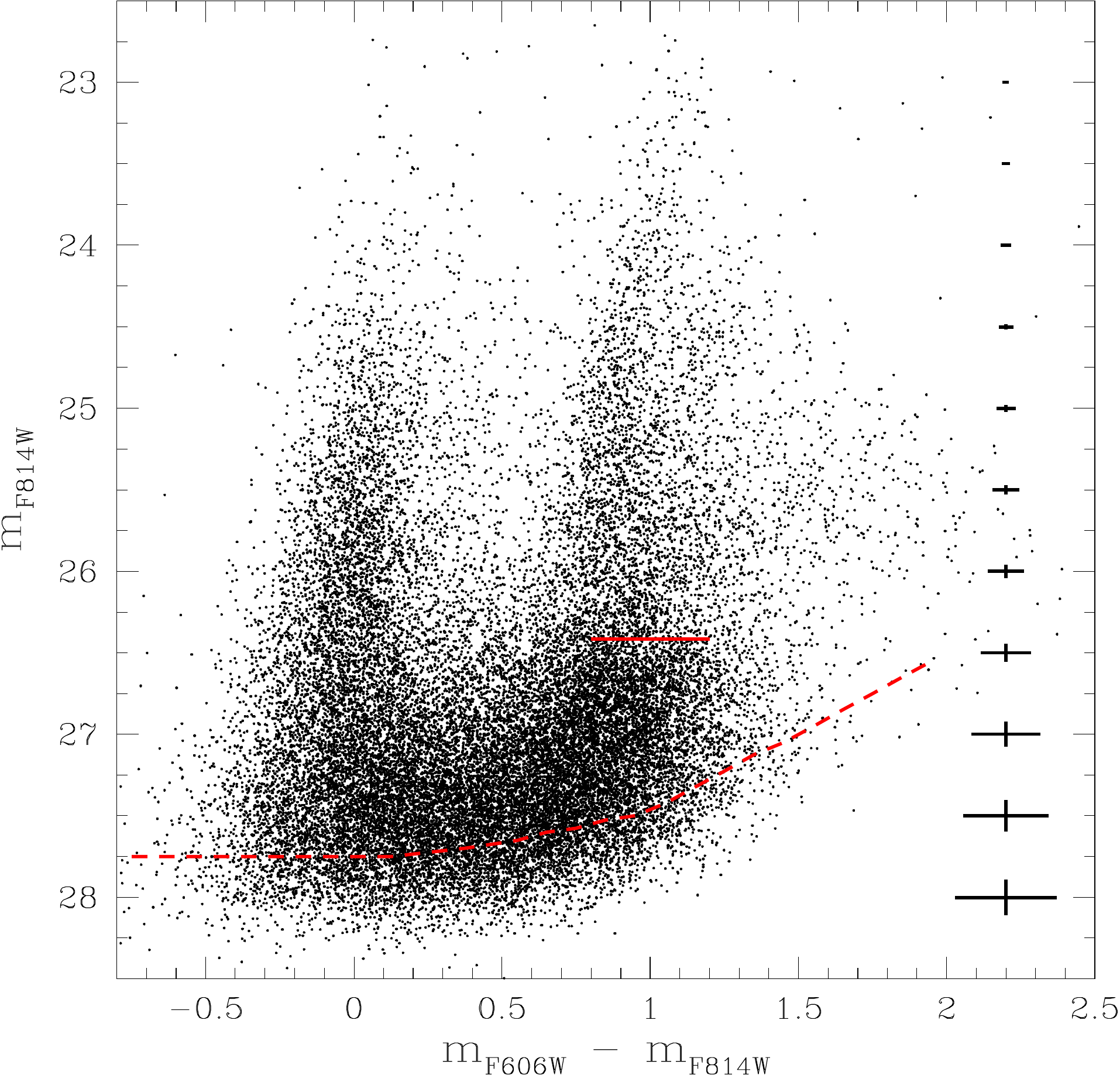}
\caption{Total CMD of the resolved stars in DDO~68. The data are corrected for foreground extinction but not for possible extinction internal to the galaxy. The photometric errors on the right are those from the artificial star tests (1 $\sigma$ of the output-input distribution) assuming that the errors are uncorrelated, and that $V-I=1$. The dashed curve provides an estimate of the 50\% completeness level. The horizontal red segment indicates the magnitude of the TRGB as computed within this work.}
\label{cmd_all}
\end{figure*}
\begin{figure*}
\includegraphics[width=\linewidth]{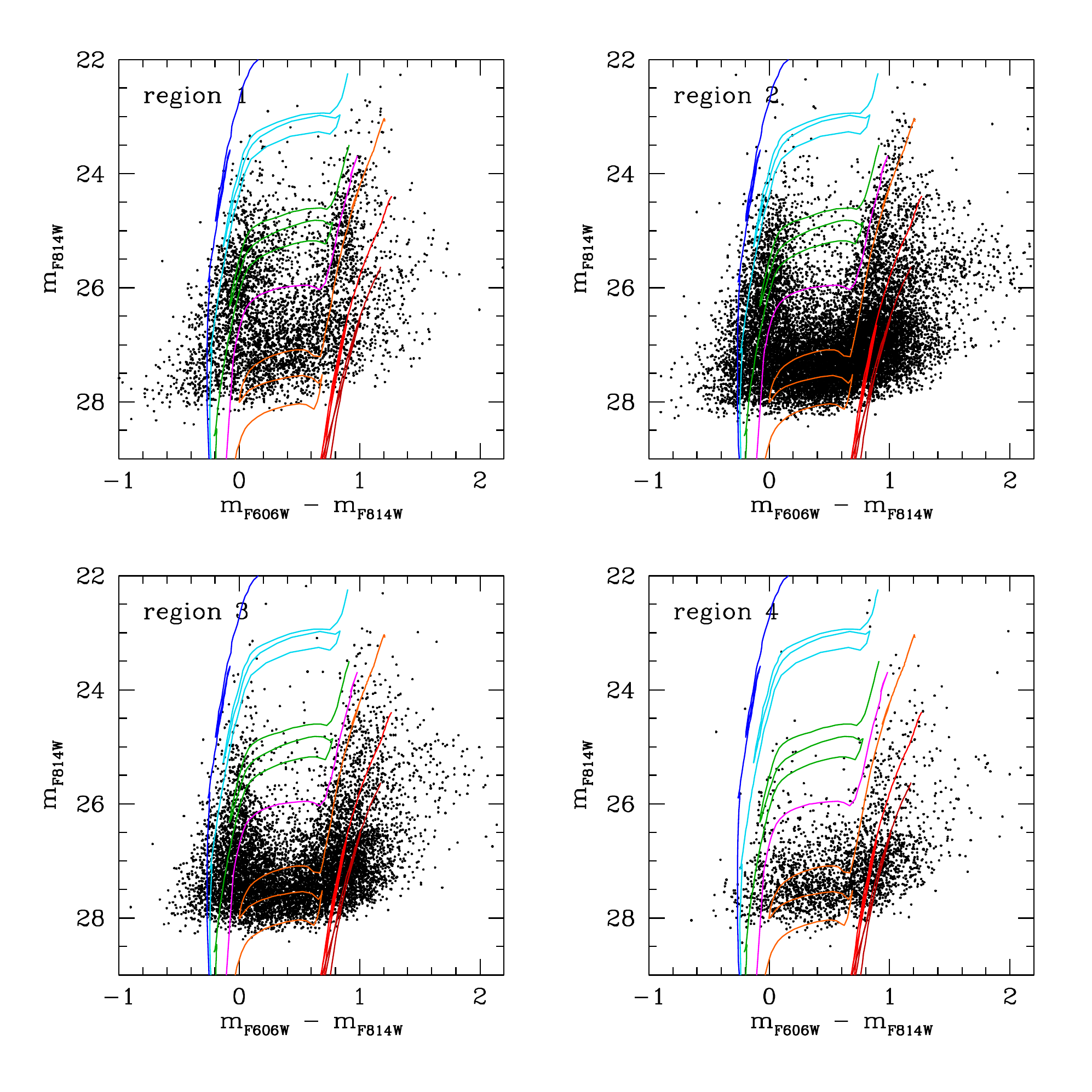}
\caption{CMDs for the four regions of the galaxy (the region numbering goes from inside out, so Region 1 is the most internal one). The isochrones superimposed (from left to right) correspond to the following ages: 10~Myr (blue), 20~Myr (cyan), 50~Myr (green), 100~Myr (magenta), 200~Myr (orange), 2~Gyr (red) and 13~Gyr (dark red), and are plotted assuming $(m-M)_0=30.51$ and $E(B-V) = 0.018$.} 
\label{4cmd}
\end{figure*}
Figure \ref{cmd_all} shows the total $\mathrm{m_{F814W}}$ vs $\mathrm{m_{F606W}-m_{F814W}}$ CMD of DDO~68 after applying the selection cuts described in Section \ref{obs}. An estimate of the 50\% completeness limit as derived from the artificial star experiments is shown in the figure, as well as the location of the TRGB as derived in Section \ref{tip_dist}.  Also shown is the average size of the photometric errors in each magnitude bin. These are the errors resulting from the artificial star tests.
The CMD shows a blue plume at $\mathrm{m_{F814W}}\lesssim 26.7$ and $\mathrm{m_{F606W}-m_{F814W}}\simeq 0$, which is the locus of young ($\lesssim 10$~Myr) main sequence (MS) stars and blue supergiants (BSGs) (evolved stars at the hot edge of their core He-burning phase), and is typical of star-forming systems. At $\mathrm{m_{F606W}-m_{F814W}} \simeq 0.9$~mag, the red plume is populated by a mix of RGB, He-burning stars near the red edge of the loop and early asymptotic giant branch (AGB) stars of intermediate mass, implying ages from $\sim 20$~Myr up to several Gyr. At intermediate colors, the CMD samples the blue loop (BL) phase of intermediate mass stars, with ages between $\sim 100$~Myr and $\sim 300$~Myr (the oldest and faintest BL sampled by our CMD). The very red objects with $25 < \mathrm{m_{F814W}} < 26$ and $\mathrm{m_{F606W}-m_{F814W}} \gtrsim 1.2$ are thermally pulsing asymptotic giant branch (TP-AGB) and Carbon stars with ages from $\sim 300$~Myr to $\sim 2$~Gyr. More importantly, the presence of a well-populated RGB below $I \simeq 26.4$ implies a population of stars at least $1-2$~Gyr old, and possibly as old as the Hubble time. The age-metallicity degeneracy affecting RGB stars normally prevents us from claiming that the redder ones are stars as old as 13~Gyr. However, the extremely low metallicity of DDO~68 makes this claim rather plausible.

In order to account for gradients in the stellar populations, we constructed separate CMDs for the 4 regions selected through isophotal contours (see Fig. \ref{map}). The CMDs are shown in Fig. \ref{4cmd} with superimposed the Padova PARSEC isochrones \citep{Bressan2012} shifted to a distance of 12.65 Mpc (Section \ref{tip_dist}) and with a foreground Galactic extinction $E(B-V)=0.018$ \citep{Schlegel1998} $-$ notice that in the PARSEC isochrones the thermally pulsing AGB phase is not included. These CMDs clearly show a gradient in the stellar populations, with younger stars detected predominantly in the most central regions. However, the comparison with the isochrones reveals a significant population of stars with ages of $\sim 10$~Myr or younger in Regions 2 and 3, in agreement with the presence there of a large number of H~II regions (see Section \ref{hsec}). The old ($> 1$~Gyr) RGB population is well detected everywhere but in Region 1. A more detailed analysis of the spatial distribution of the different stellar populations will be discussed in Section \ref{pop_distr}. 

\section{Distance Determination}
\label{tip_dist}
\begin{figure}
\includegraphics[width=\columnwidth]{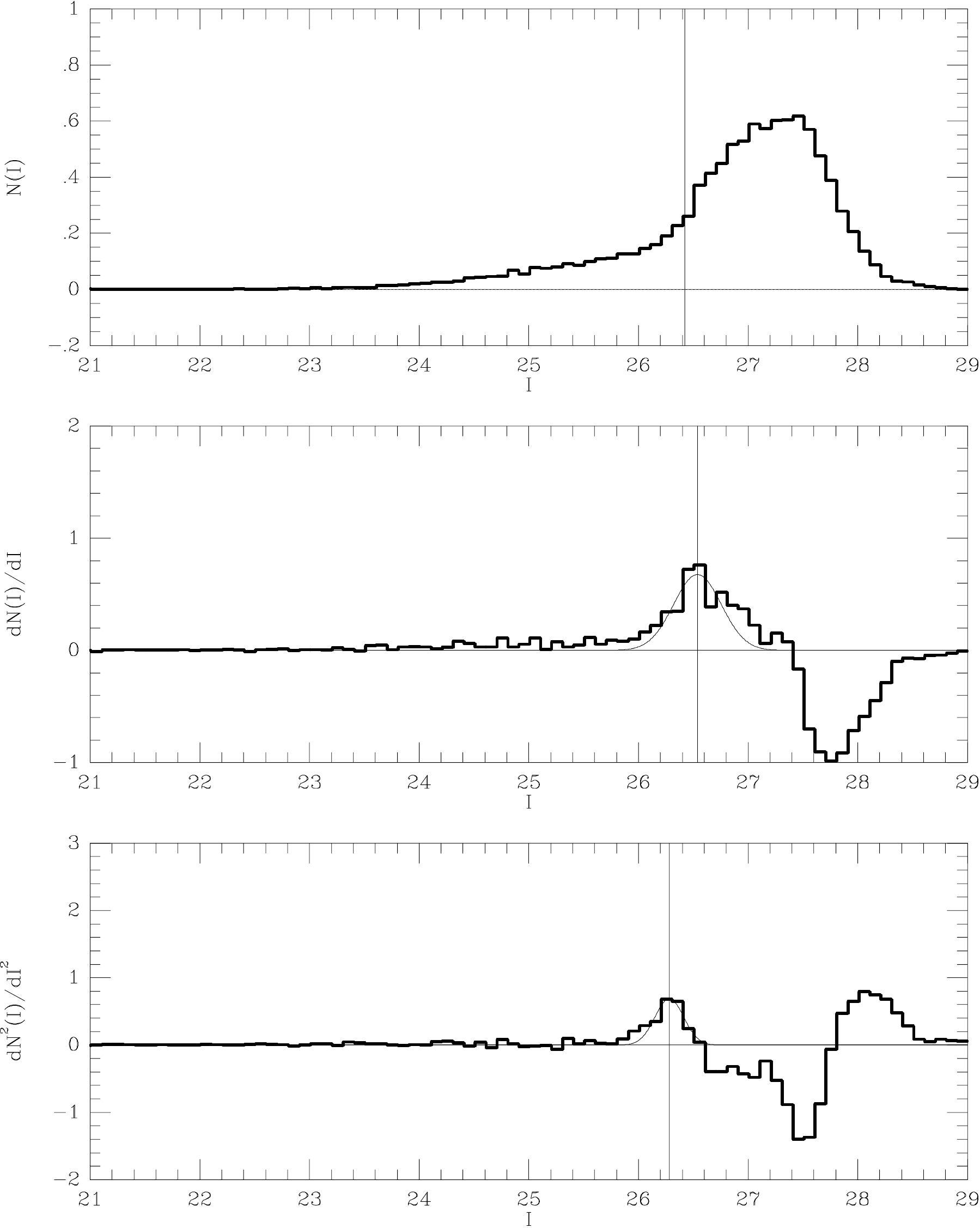}
\caption{$I$-band LF (top) and its first (middle) and second (bottom) order derivatives for stars with $V-I$ in the range $0.5-1.3$. Magnitudes are corrected for extinction. The vertical lines indicate the position of the TRGB as derived in Section \ref{tip_dist}.  The normalization of all the vertical scales is arbitrary.}
\label{tip}
\end{figure}
\begin{figure}
\centering
\includegraphics[width=0.98\columnwidth]{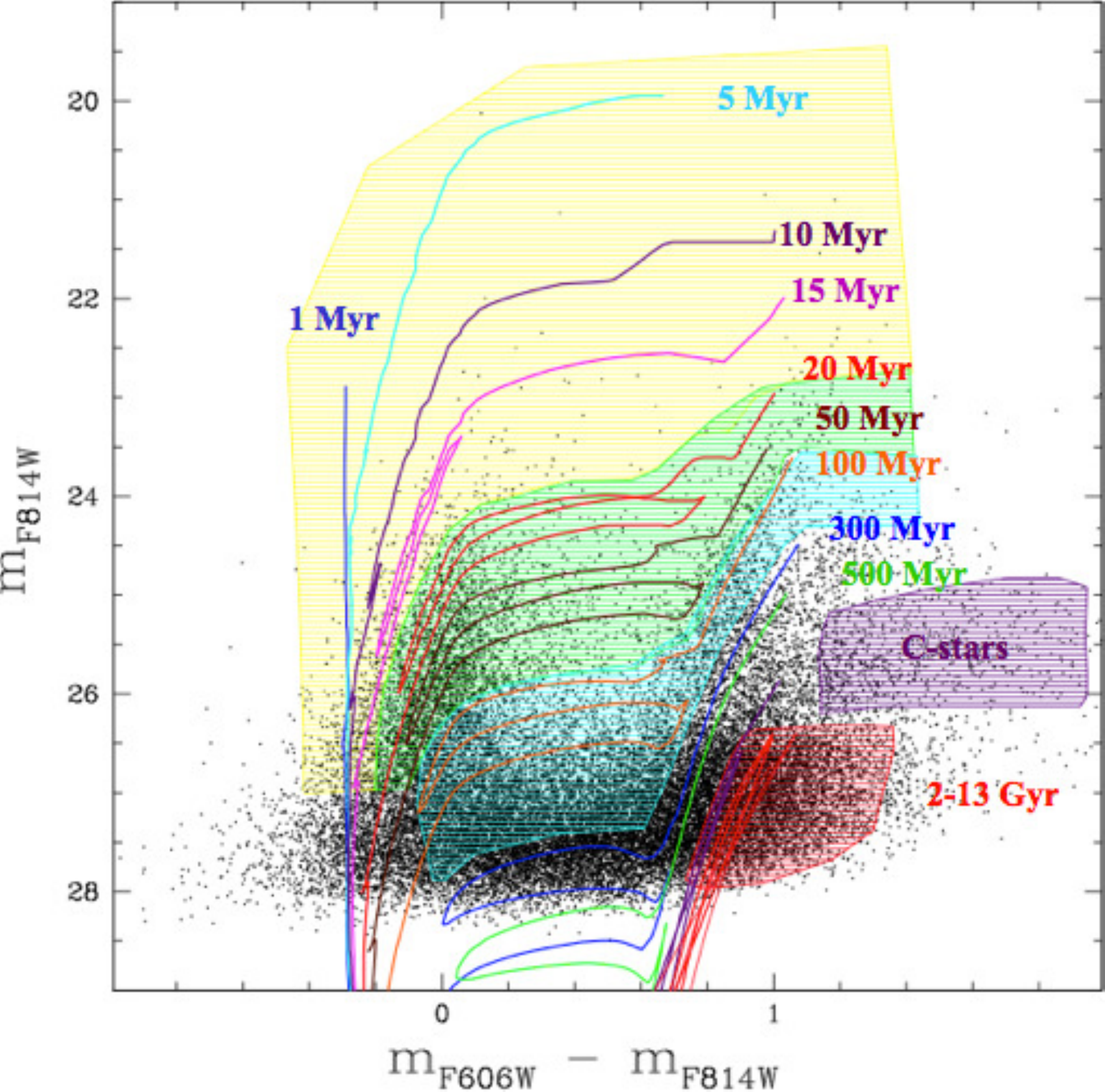}
\caption{Selection of different age intervals identified through the Z=0.0004 stellar isochrones \citep{Bressan2012} shown and labeled in the image.}
\label{pop_cmd}
\end{figure}

We derived the distance of DDO~68 using the magnitude of the TRGB. We show in Figure \ref{tip} the $I-$band luminosity function (LF) of stars with color \mbox{$0.5<V-I<1.3$} in our catalog (the photometry was converted to the Johnson-Cousins system through the procedure outlined in \citealt{Sirianni2005} in order to apply the described method).

We visually identified the TRGB as a steep increase in the LF at $I = 26.416$. There, RGB stars start to contribute to the LF, while at brighter magnitudes only red supergiants and AGB stars contribute. The dip in the LF  fainter than $I \simeq 27.5$ is due to incompleteness effects. The TRGB magnitude was determined with the approach described by \cite{Cioni2000}, that uses the peaks in the first and second order derivatives of the LF to better identify the position of the TRGB. These peaks are shown in the middle and bottom panels of Figure \ref{tip}. Once photometric errors and binning effects are accounted for, the first and second order derivative peaks provide a TRGB magnitude of  $I_{TRGB,1} = 26.424$ and $I_{TRGB,2} = 26.408$, respectively. The systematic error of this procedure was estimated by \cite{Cioni2000} to be $\Delta I_{TRGB} = \pm0.02$. The additional systematic errors, due to uncertainties in the photometric zeropoints and conversions, and in aperture corrections \citep{Sirianni2005}, is $\Delta I_{TRGB} = \pm0.023$. The random error on $\Delta I_{TRGB}$ from the finite number of stars, estimated using bootstrap techniques, was found to be $\Delta I_{TRGB} = \pm 0.057$. Addition of these errors in quadrature yields a total (i.e. random plus systematic) error of $\pm 0.068$~mag. After correcting for an $I-$band foreground extinction towards DDO~68 of $A_I = 0.036$ [E($B-V$) = 0.018], we derive the final TRGB magnitude estimate of $I_{TRGB,0} = 26.380 \pm 0.068$. To derive the distance modulus, we need to compare the observed TRGB magnitude with the absolute one, which was calibrated as a function of metallicity by, e.g., \cite{Bellazzini2004}. At the metallicity of I Zw 18, a comparable metallicity to that of DDO~68, this is estimated at $M_{I,TRGB} = -4.03\pm 0.10$ \citep{Aloisi2007} which implies a distance modulus of $(m-M)_0 = 30.41 \pm 0.12$~mag, i.e., $D = 12.08 \pm 0.67$~Mpc, in agreement with the value obtained by \cite{Tikhonov2014} from the same data.

We should however consider that the location of the TRGB is likely affected also by a significant blending of multiple stars in these crowded regions. The effect of blending is always to let the TRGB appear brighter. Indeed, when fitting the observed CMDs with synthetic ones to derive the SFH (see Section \ref{sec_sfh}), we always found that all the evolutionary sequences are much better reproduced if the distance modulus is slightly larger, although within the uncertainties. We therefore conclude that the distance modulus of DDO~68 is $(m-M)_0 = 30.51$, and $D=12.65$~Mpc, in agreement with the independent estimate by \cite{Cannon2014}. From now on we therefore adopt the latter distance for all applications.\\
\begin{figure*}
\centering
\includegraphics[width=2\columnwidth]{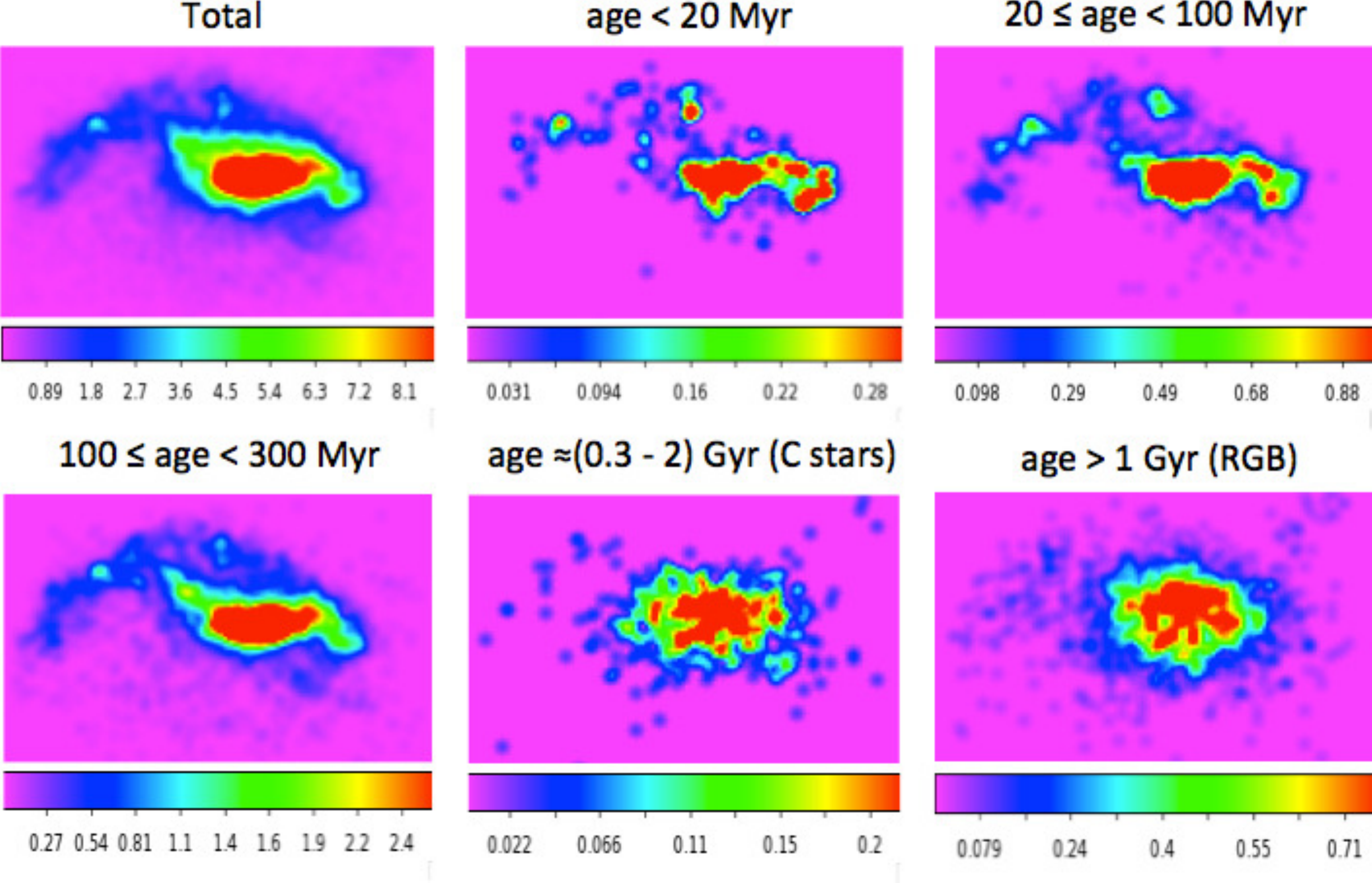}
\caption{Density maps of stars observed in different evolutionary phases within DDO~68, as selected through their position in the CMD (see Fig. \ref{pop_cmd}). The scale is in number of stars/arcsec$^2$.}
\label{pop_maps}
\end{figure*}

\section{Stellar populations in DDO~68}
\subsection{Spacial distribution} \label{pop_distr}
\begin{figure*}
\includegraphics[width=\columnwidth]{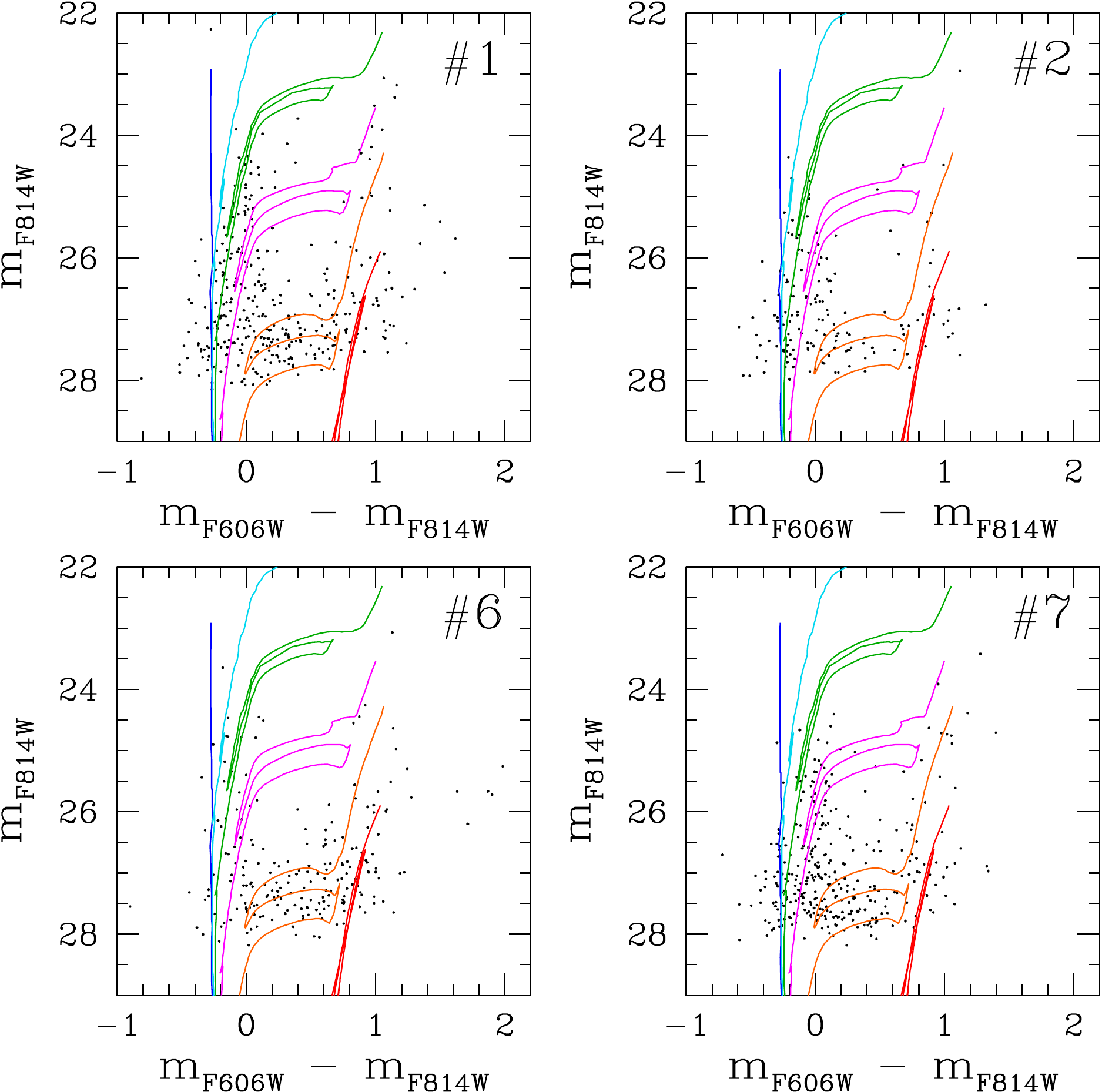}
\hspace{0.1cm}
\includegraphics[width=\columnwidth]{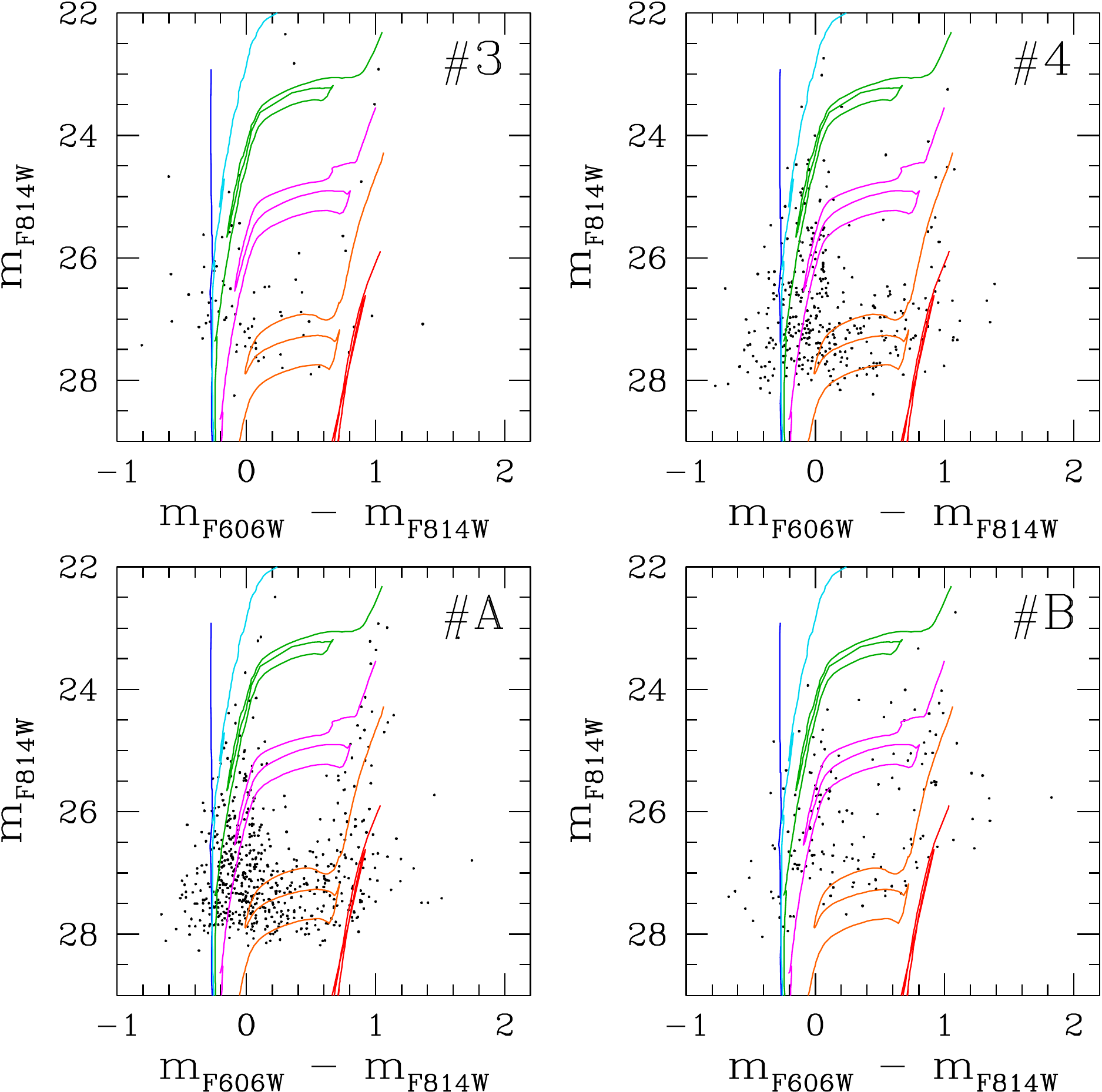}
\caption{CMDs of the H II regions in DDO~68. The isochrones superimposed (from left to right) correspond to the following ages: 1~Myr (blue), 10~Myr (cyan), 20~Myr (green), 50~Myr (magenta), 200~Myr (orange) and 2~Gyr (red). The numbers follow the labeling by \cite{Pustilnik2005} and \cite{Izotov2009} except for the last two regions (A and B) that they don't take into account in their analysis.}
\label{HIIcmds}
\end{figure*}
As a first qualitative analysis, we looked at how the different stellar populations are distributed over the galaxy. To this purpose, we used the PARSEC \citep{Bressan2012} Z=0.0004 stellar isochrones (corresponding to the metallicity of DDO~68's H~II regions, \citealt{Izotov2009}) to identify on the CMD a few regions corresponding to different age intervals (see Fig. \ref{pop_cmd}): age~$< 20$~Myr, $20 \lesssim$~age $< 100$~Myr, $100\lesssim$~age~$<300$~Myr, and age $>1$~Gyr (the RGB phase); we also identified the region of the Carbon-stars (not covered by the adopted isochrones), with typical ages between $\sim 0.3$ and 2~Gyr. The spatial distribution of the stars in the different age intervals, as well as the global distribution for the total number of stars in the CMD, are shown in the star count density maps of Fig. \ref{pop_maps}. To the purpose of better highlighting some particular features, the images were smoothed and arbitrarily scaled. From an inspection of the spatial maps in Fig. \ref{pop_maps}, we immediately notice an important difference with the typical behavior of the stellar populations in other star-forming dwarfs, where the youngest stellar populations are preferentially concentrated toward the most central galaxy regions: in the case of DDO~68, in fact, stars younger than $\sim 20$~Myr are also found at relatively large galactocentric distances, and in particular they appear to follow the arc-shaped structure well visible in the total spatial map. This is consistent with the presence there of numerous H~II regions, as highlighted in Fig. \ref{map}, which trace stars typically younger than $\sim 10$~Myr. A similar behavior was also found in the starburst dwarf NGC~4449, for which the presence of young stars at high galactocentric distances is likely due to an accretion event \citep{Annibali2012}. As we consider increasingly older stellar populations, the distribution becomes more and more homogeneous, until the old ($> 1$~Gyr) RGB population traces really well the shape of the whole galaxy. Notice that the paucity of RGB stars in the center of DDO~68 is an artifact due to the strong incompleteness affecting the most crowded Region 1, and does not reflect the intrinsic distribution of the old stellar populations. Interestingly, we can see the disturbed arc-shaped Tail in all the age intervals, meaning that this structure is not made only of young stars, as the distribution of the H$\alpha$ emission might have suggested. The presence of stars of all ages in the arc-shaped structure suggests that this is a possible dwarf galaxy in interaction with the central body of DDO~68, since stars older than $\sim 1$~Gyr wouldn't be able to remain in the form of a gravitationally unstable arc. Indeed, in a recent work, \cite{Tikhonov2014} suggested that DDO~68 consists of two distinct objects: a more massive central galaxy, DDO~68~A, traced by the spatial distribution of the bulk of the old stars, and a dwarf galaxy in interaction with DDO~68~A, namely DDO~68~B, corresponding to the arc-shaped structure, an hypothesis that we find very interesting and worth of further investigation.

In the RGB spatial map we also recognize a small concentration of stars toward the upper right edge of the ACS chip, noticed also by \cite{Tikhonov2014}; new results on this feature will be presented in a forthcoming paper based on new LBT data with a much larger field of view \citep{Annibali2016}.

\subsection{H II regions} \label{hsec}
The F658N narrow band filter allows us to study how the H$\alpha$ emission is located within the galaxy and to identify the very recent concentrations of SF. In fact, the H$\alpha$ recombination line is one of the most widely used tracers of recent SF, since it comes from the recombination of gas ionized by photons of massive stars ($>20$~M$_{\odot}$) and thus is expected to be observed over the typical lifetimes of massive stars ($< 10$~Myr) \citep{Leitherer1999}. Thus, the ionized gas emission traces the youngest SF and provides an important piece of information on the SFH and on the dynamical interactions of the galaxy. A quite interesting and uncommon feature is that the \mbox{H II} regions in DDO~68 are placed mostly in the outer parts (see Figs. \ref{3col} and \ref{map}) suggesting some external effect to trigger the SF. As already discussed, this could be the presence of another object which has recently interacted with DDO~68. We selected 8 H~II regions based on the gas distribution, and visually analyzed the CMDs of their resolved stars (Fig. \ref{HIIcmds}) finding that they are predominantly populated by young blue-plume stars. A comparison with stellar isochrones indicates that stars younger than $\sim 10-20$~Myr are found in these regions, in agreement with the presence of ionized gas. The regions in Fig. \ref{HIIcmds} were labeled following \cite{Pustilnik2005} and \cite{Izotov2009}.\\

\section{Star Formation History} \label{sec_sfh}
The SFH of DDO~68 was derived by comparing the observed CMDs with synthetic ones, following the method originally devised by \cite{Tosi1991}. The synthetic CMDs were constructed as linear combinations of ``basis functions'' (BFs), i.e contiguous star formation episodes whose combination spans the whole Hubble time. Observed and synthetic CMDs were compared through a statistical analysis; to this purpose we adopted two independent procedures: one originally developed by R. P. van der Marel and described in \cite{Grocholski2012}, based on the code SFHMATRIX, and the other devised by M. Cignoni and described in \cite{Cignoni2015}, based on SFERA (Star Formation Evolution Recovery Algorithm).

We consider it useful to compare the SFHs derived with the two different approaches to estimate which results are robust and which are more uncertain or even artifacts of the chosen minimization algorithms.

\subsection{Basis Functions}
The construction of the BFs is based on the combination between stellar evolution models and observational uncertainties (e.g. photometric errors, incompleteness) of the examined galaxy region. We created the BFs starting from the latest (V.1.2S) Padova PARSEC isochrones \citep{Bressan2012}, which include the pre-main sequence evolutionary phase (while the TP-AGB phase is still not included). 
We adopted a Kroupa IMF \citep{Kroupa2001} from 0.08 to 120~M$_{\odot}$ and we assumed a 30\% fraction of binary stars, whose masses were extracted from the same IMF of the ``main'' population.

The BFs were computed up to an age of $\sim 13.5$~Gyr with a logarithmic time duration of $\Delta$log(age) = 0.25, except for ages younger than 1 Myr which were grouped into a single time bin. The logarithmic step is chosen to take into account the lower time resolution as a function of the look-back time. The star formation rate (SFR) is assumed to be constant within each episode.

The ``theoretical'' BFs were convolved with the characteristics of the data, i.e. distance, foreground and internal extinction, as well as the completeness and photometric errors derived from the artificial star tests. Absolute magnitudes were converted into apparent ones applying a foreground reddening of $E(B-V) = 0.018$ from the NASA/IPAC Extragalactic Database\footnote{The NASA/IPAC Extragalactic Database is operated by the Jet Propulsion Laboratory, California Institute of Technology, under contract with the National Aeronautics and Space Administration. \url{https://ned.ipac.caltech.edu}}, and a distance modulus allowed to vary by 1 mag around the value $(m-M)_0 = 30.41$~mag derived from the TRGB (see Section \ref{tip_dist}). Eventually, the best value to fit all the evolutionary sequences and LFs turned out to be $(m-M)_0 = 30.51$. 

To incorporate observational effects and systematic errors (due to e.g. photometric blends) we assigned  photometric errors to the synthetic stars using the difference between output and input magnitudes of the artificial stars (see Figure \ref{complerr}). To make sure that all the stellar evolutionary phases are well populated  in spite of the incompleteness, the synthetic CMDs are generated with a very large number of stars. The comparison with the observational CMD is however performed considering the number of stars actually present in the latter. The comparison is done on the Hess diagram,  i.e., considering the density of points in the CMD. To this purpose, the BFs, as well as the observed CMD, are pixelated into a grid of $n$ color-bins and $m$ magnitude-bins.

We have used the two different procedures SFERA and SFHMATRIX to identify the weighted combination of BFs that best reproduces the observational CMDs and thus the best SFH. In the following we describe the two statistical codes used for the derivation of the SFH.

\subsection{The \emph{SFERA} procedure}
To explore the wide parameter space involved in the derivation of the SFHs, SFERA combines a genetic algorithm with a local search routine, as described in detail by \cite{Cignoni2015}.

The metallicity is treated as a free parameter. We don't assume an age-metallicity relation nor the same metallicity for all the stars; however, since we have the information on the current metallicity (corresponding to $Z \sim 0.0004$) from the H~II regions, we allow the  metallicity to vary in the range [0.0002 $-$ 0.002]. For all regions, the metallicity inferred by SFERA in the youngest bins is between $Z\sim 0.00035$ and $Z\sim 0.0005$, consistent with the observed values from the H~II regions.

We end up with a library of $j \times k$ 2D histograms, BF$_{m,n}(j,k)$, that can be linearly combined to express any observed CMD, as in Equation (\ref{sfera1}). The coefficients $w(j,k)$ are the weights of each of the BF$_{m,n}(j,k)$, and represent the SFR at the time step $j$ and metallicity step $k$. The sum over $j$ and $k$ of $w(j,k) \times$ BF$_{m,n}(j,k)$ gives the total star counts $N_{m,n}$ in the CMD bin $(m,n)$:
\begin{equation}
N_{n,m}=\sum_j \sum_k w(j,k) \times \mathrm{BF}_{m,n}(j,k)
\label{sfera1}
\end{equation}
The minimization of the residuals between data and models is implemented in SFERA taking into account the low number counts in some CMD cells, so we follow a Poissonian statistics, looking for the combination of BFs that minimizes a likelihood distance between model and data. This corresponds to the most likely SFH for these data, with an uncertainty given by the sum in quadrature of a statistical error (computed through a data bootstrap) and a systematic error (obtained by re-deriving the SFH with different age and CMD binnings, even if this tends to smooth out the SFH).
The Poisson based likelihood function we chose is:
\begin{eqnarray}  
\chi_P= \sum_{i=1}^{N
    bin} obs_{i}\ln\frac{obs_{i}}{mod_{i}}-obs_{i}+mod_{i}
\label{pois1} 
\end{eqnarray}
where $mod_{i}$ and $obs_{i}$ are the model and the data histogram in the $i-$bin.
This likelihood is minimized with a hybrid-genetic algorithm, which combines a classical genetic algorithm (Pikaia\footnote{Routine developed at the High Altitude Observatory and available in the public domain: \url{http://www.hao.ucar.edu/modeling/pikaia.php.}}) with a local search (Simulated Annealing). This allows to take advantage of the exploration ability of the former, which scans the parameters space in more points simultaneously and thus isn't too sensitive to the initial conditions, and the capability of the latter of local exploration of the space and faster convergence. The proposed algorithm alternates two phases: the search for a quasi-global solution by the genetic algorithm and the local search by the other one, which increases the solution accuracy.

\begin{figure}
\centering
\includegraphics[trim=1.5cm 0 0 -0.1cm, clip, width=0.93\columnwidth]{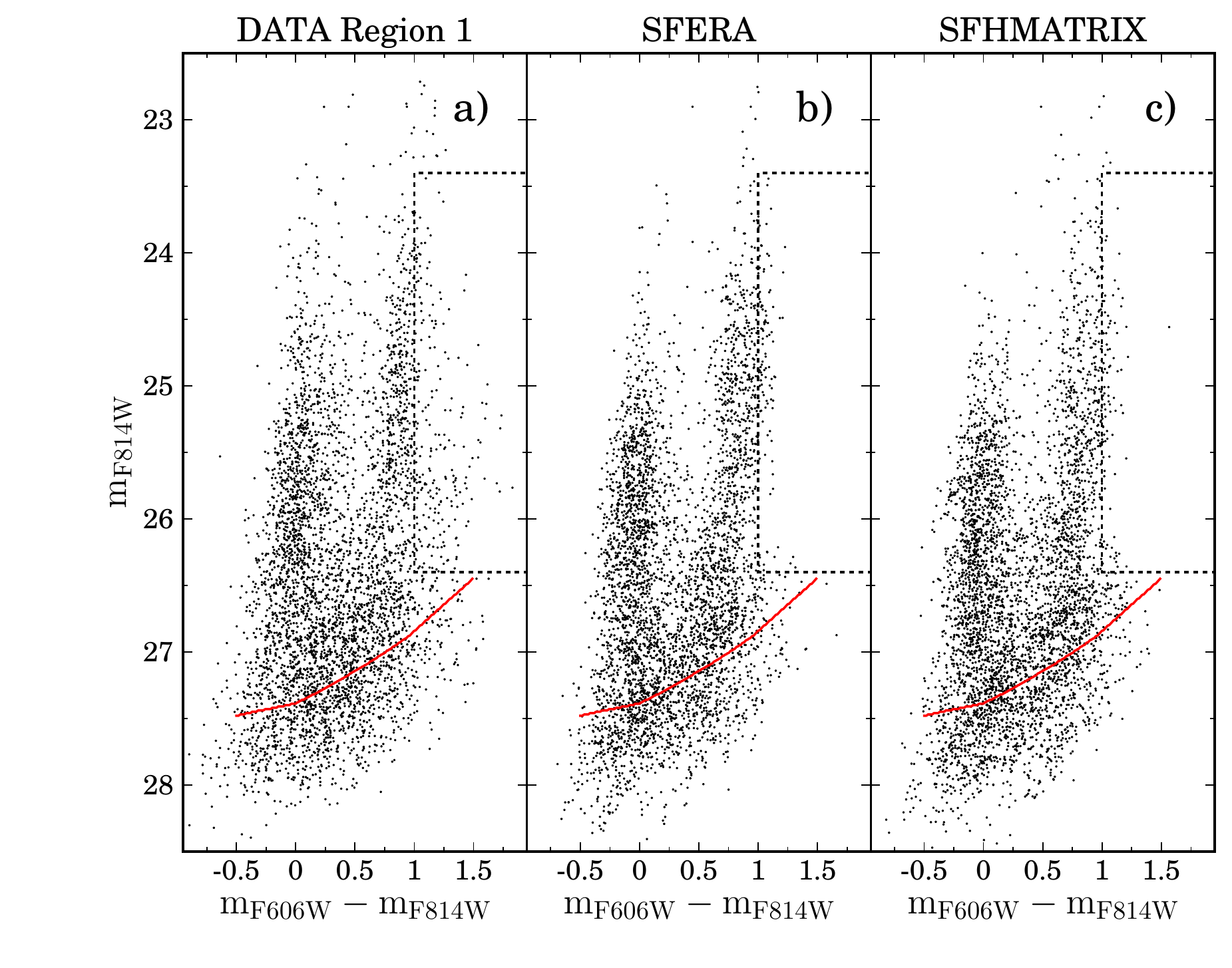}
\includegraphics[trim=0 0.4cm 0 1cm, clip, width=\columnwidth]{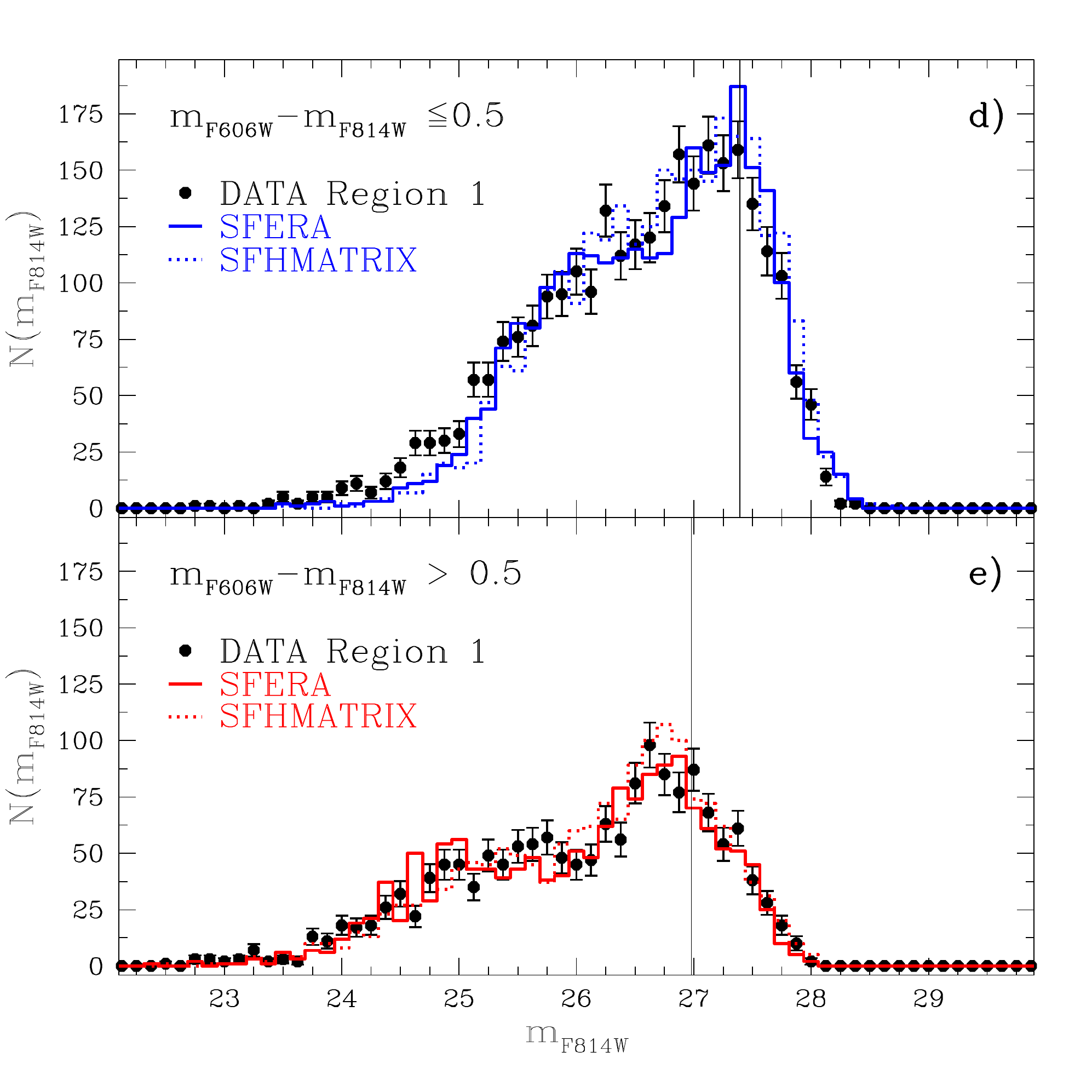}
\includegraphics[trim=0 0 0 0.5cm, clip, width=\columnwidth]{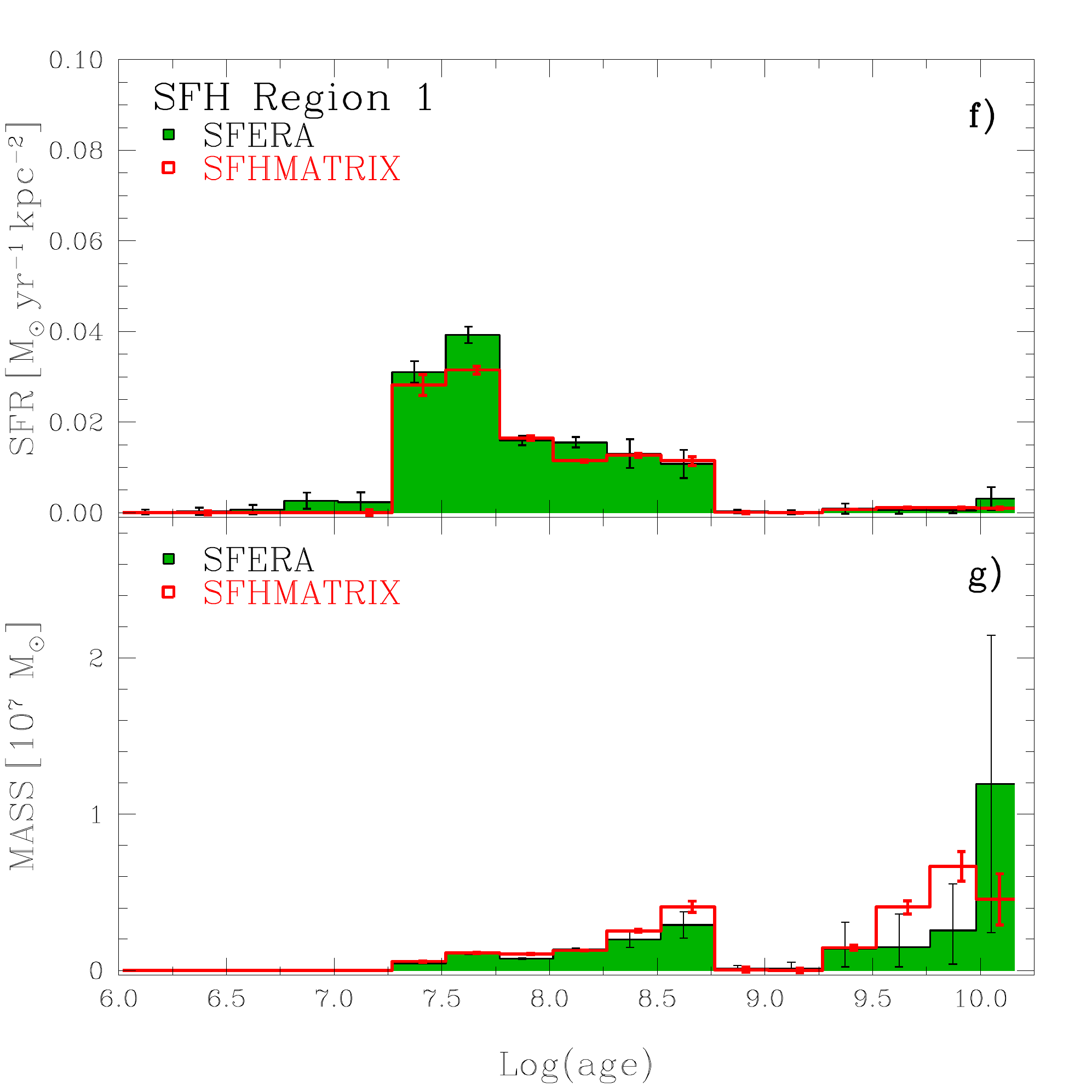}
{\color{white} \caption{}}
\label{region1}
\end{figure}
\addtocounter{figure}{-1}
\begin{figure}
\caption{Results of the synthetic CMD method for Region~1. \emph{Top panels}: a) observed CMD for Region~1, b) best-fit synthetic CMD with SFERA and c) best-fit synthetic CMD with SFHMATRIX (both obtained considering only the region with completeness above 30\% indicated by the red solid curve). The dotted box shows the TP-AGB region that was excluded from the minimization. \emph{Middle panels}: LF in the blue (d) and red (e) for the observed (black points with Poissonian error) and synthetic CMDs (solid line for SFERA, dotted line for SFHMATRIX). The vertical lines indicate the average faintest magnitudes that were considered for the fit. \emph{Bottom panels}: best-fit SFH (f) normalized over the area of the region and recovered stellar mass (g) with SFERA (green solid histograms) and SFHMATRIX (red histograms). Notice that the error bars coming out from SFHMATRIX have to be considered as lower limits to the real uncertainties of the solution, since only statistical errors are included and the metallicity is fixed.\\}
\end{figure}
\subsection{The \emph{SFHMATRIX} procedure}
The SFHMATRIX code solves a non-negative least-squares (NNLS) matrix problem to identify the SFH that best reproduces the Hess diagram of the observational CMD in a $\chi^2$ sense. The minimization problem is addressed as the solution of a matrix equation:
\begin{equation}
\sum_j w(j) \times \mathrm{BF}_{m,n}(j) = \rho_{m,n} \pm \Delta\rho_{m,n}
\label{somma2}
\end{equation}
where $\rho_{m,n}$ is the star density in the observed CMD, $\Delta\rho_{m,n}$ is the corresponding Poisson uncertainty, and $m$ and $n$ are the Hess diagram pixels; the density $\rho_{m,n}$ is given by the integer number of stars $L_{m,n}\geq 0$ that is detected in Hess diagram pixel $(m,n)$, divided by the area of that pixel. The adopted Poisson error on the detected number of stars is $\max(1,\sqrt{L_{m,n}})$, hence:
\begin{equation}
\frac{\Delta\rho_{m,n}}{\rho_{m,n}}=\frac{\max(1,\sqrt{L_{m,n}})}{L_{m,n}}.
\end{equation}
$\mathrm{BF}_{m,n}(j)$ is the density of stars in the BF corresponding to the $j-th$ time step in the same pixel of the Hess diagram, weighted for the coefficient $w(j)$ (see \citealt{Grocholski2012} for the details). Since SFHMATRIX is primarily used as a cross$-$check and validation of the SFERA results, we use BFs of fixed low metallicity ($Z \sim 0.0004$).

The errors on the SFH are estimated by drawing many Monte Carlo realizations from the best-fit SFH and analyzing them with the same procedure used for the real data.

\cite{Mighell1999} demonstrated that a $\chi^2$ minimization is generally biased when data are Poisson distributed; in particular, assuming the uncertainty on the data $\Delta\rho_{m,n}=\max(1,\sqrt{L_{m,n}})$ leads to an underestimation of the true mean of the Poisson distribution. Given this uncertainty, we use the $\chi^2$ approach only as a comparison of the SFERA results. In the next sections, we explore and compare in details the results from the two different adopted procedures and we will come back on this bias in Section \ref{sect_region4}.

\subsection{\textsc{Region 1}}
Panel a) of Fig. \ref{region1} shows the observed CMD of the most crowded Region~1 of DDO~68, together with the synthetic realizations inferred from the best-fit solutions of the SFERA (panel b) and SFHMATRIX (panel c) codes. We excluded from the minimization the cells of the CMD with a completeness below 30\% where the uncertainties become significant. We used a lower completeness threshold with respect to the other regions in order to reach at least the TRGB and have a strong constraint on the CMD. We run several tests using a different completeness limit (up to 50\%) and this yielded results that are qualitatively similar. We choose to keep the results with a completeness down to 30\% to include the RGB stars, since they are the only signature of epochs older than $1-2$ Gyr ago; we do take into account the larger uncertainty due to this choice and we are confident that this doesn't affect the main conclusions of the paper.

When using SFHMATRIX, we created a grid with a constant cell size of 0.5 both in color and in magnitude: given the low number of stars in Region~1, this cell size allows to minimize the statistical fluctuations without loosing information on the CMD. For the same reason, we treated the box at $0.5 < \mathrm{m_{F606W}-m_{F814W}} < 1.3$ and $26.4 < \mathrm{m_{F814W}} < 29$ (i.e. the RGB) as a single pixel, while we excluded from the minimization the box at $1 < \mathrm{m_{F606W}-m_{F814W}} < 2$ and $23.4 < \mathrm{m_{F814W}} < 26.4$, corresponding to the TP-AGB phase which is not included in the PARSEC isochrones yet.

In SFERA, where flexibility in the construction of the grid is possible, we chose a quite large cell size (0.5 both in color and magnitude) for the brightest stars with $\mathrm{m_{F814W}} < 24$ to balance the low statistics caused by the few number of bright stars; we chose an intermediate cell size (0.1 in color, 0.2 in magnitude) for the blue and red plumes and for the lower MS; for the RGB instead we implemented a variable random binning from 1 to 16 cells in the box $0.5 < \mathrm{m_{F606W}-m_{F814W}} < 1.3$ and $26.4 < \mathrm{m_{F814W}} < 29$, that we changed at every bootstrap in order to minimize the bin dependence of the results.

To reproduce the observed CMD and LF in Region~1, as well as in all the other regions of DDO~68, we found more satisfactory to adopt a slightly higher distance modulus than that derived from the TRGB [that is $(m-M)_0 = 30.51$ instead of $(m-M)_0 = 30.41$]; this larger distance modulus, which is still inside the errors of the previous determination, implies a distance of 12.65~Mpc.

We show in Fig. \ref{region1} the solutions found by the two codes for the best CMD (b-c), LF (d-e) and SFH (f-g). It is evident that the codes are both able to well reproduce the overall morphology of the observed CMD, although with some differences. In particular, both codes tend to under-predict the number of stars in the blue LF ($\mathrm{m_{F606W}-m_{F814W}} \leq 0.5$) at $24 < \mathrm{m_{F814W}} < 25.25$, i.e. the blue supergiants. A similar problem was found for another very low-metallicity dwarf galaxy, I~Zw~18, for which we derived the SFH \citep{Annibali2013}. As already discussed in that study, possible explanations for the observed discrepancy are: i) the completeness behavior with magnitude is not properly taken into account, because of the severe crowding of this central region; ii) some of the most luminous blue stars in  the CMD are not actually single objects but blends of two or more blue stars; iii) the timescales for the evolution of massive stars in the brightest post-MS phase are not properly modeled.

It is plausible that models including additional parameters, like stellar rotation \citep{Meynet2002}, could improve the fit of the brightest objects in the CMD, but these models are not implemented yet for DDO~68's metallicity.

Panels f) and g) of Fig. \ref{region1} shows the best-fit SFHs (normalized over the area of the region) and the stellar mass assembled in the different time bins, where the solid green and the red histograms are the output of SFERA and SFHMATRIX, respectively. These SFHs are quite in agreement with each other, and clearly show an old activity, once again confirming the presence of a population of stars at least 1~Gyr old and probably as old as the Hubble time. Notice that the low uncertainty inferred for some bins with very low SF (e.g. $8.75-9.25$) could be an artifact of the bootstrap approach (see \citealt{Dolphin2013} for a detailed discussion). The mass in stars at epochs older than 1~Gyr is $\sim$ 70\% of the total stellar mass formed in this region. There is no evidence of a very young ($< 5$~Myr) SF episode, while the strongest burst (\mbox{SFR$_{\, peak} \simeq 3.9 \times 10^{-2}$ M$_{\odot}$/yr/kpc$^2$}) occurred between 20 and 60~Myr ago. 

The results for the average SFR and for the stellar mass formed in different time intervals are summarized in Table \ref{tabella}.\\

\subsection{\textsc{Region 2}}
\begin{figure}
\centering
\includegraphics[width=0.9\columnwidth]{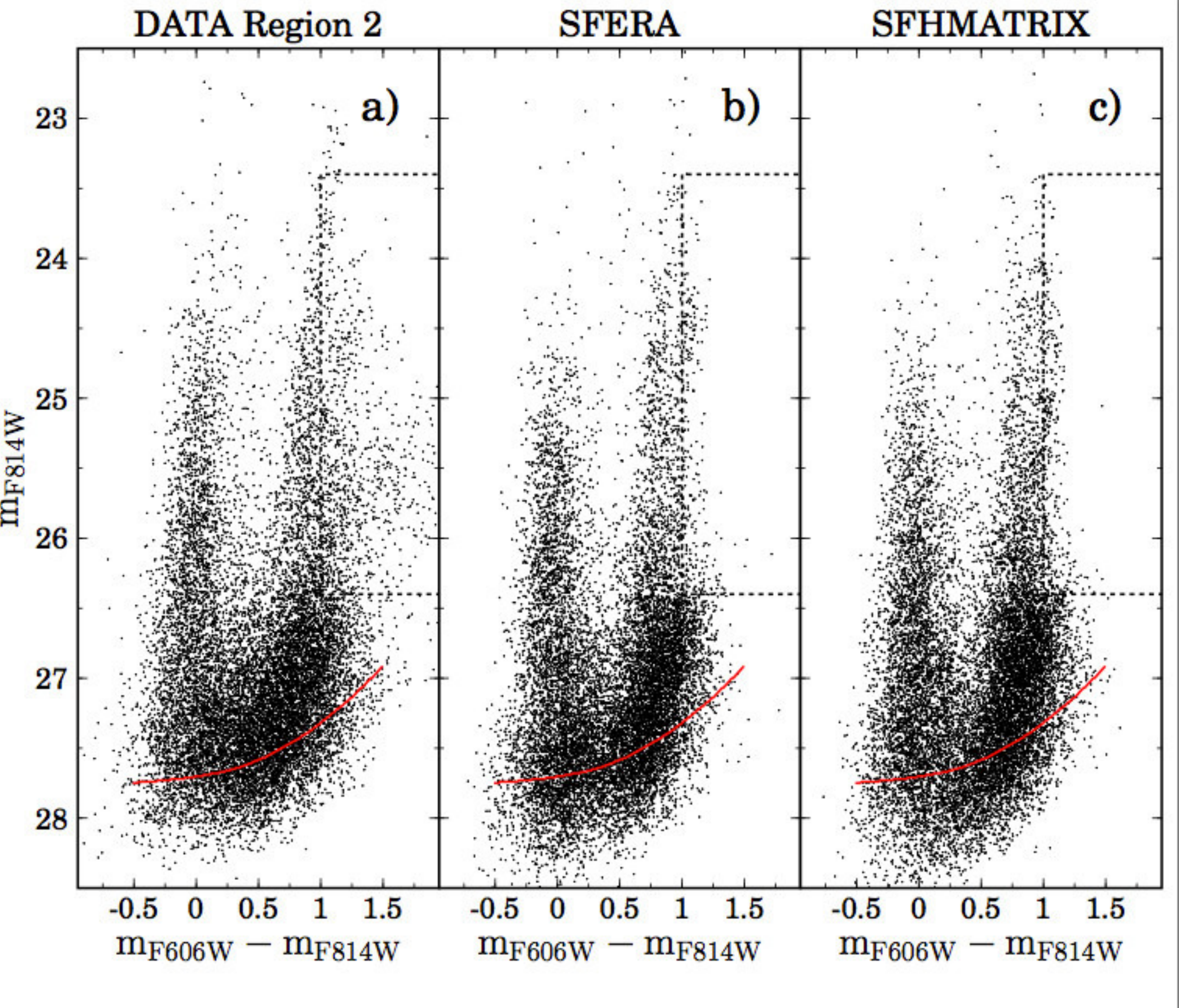}
\includegraphics[trim=0 0.4cm 0 1cm, clip, width=\columnwidth]{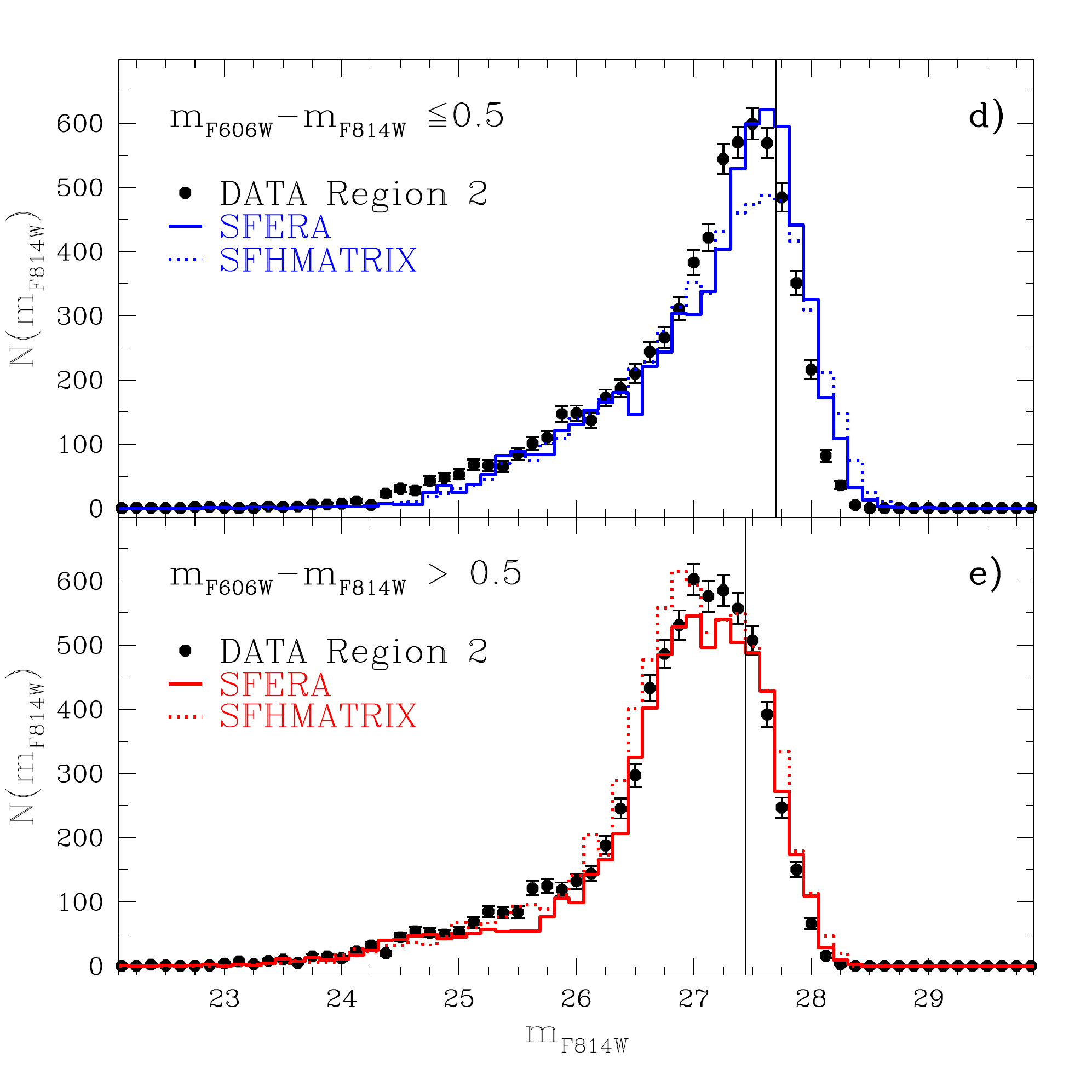}
\includegraphics[trim=0 0 0 0.5cm, clip, width=\columnwidth]{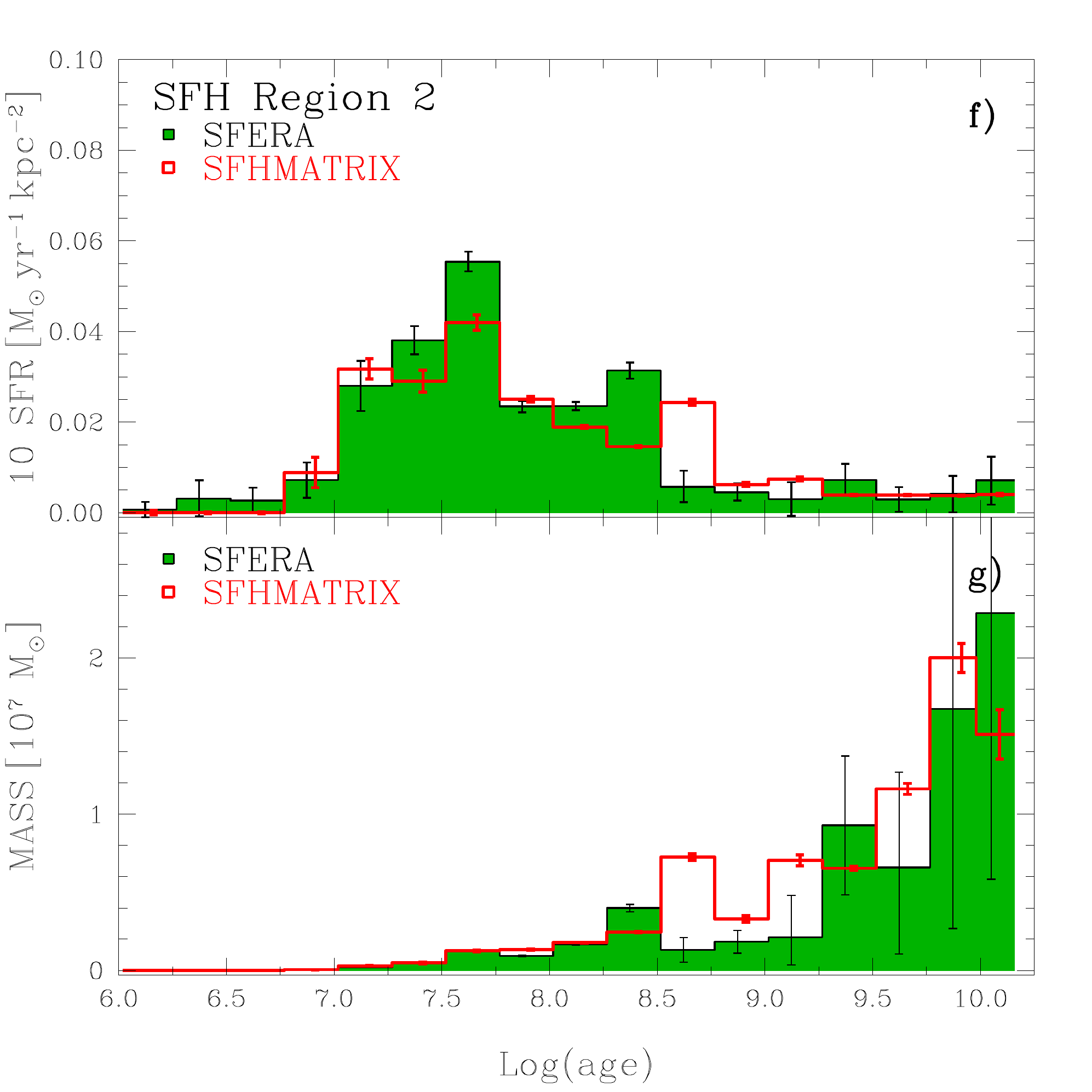}
\caption{Same as Fig. \ref{region1} but for Region 2 (the best-fit is performed in the region with completeness above 50\%). Notice that in this case the SFRs of panel f) are multiplied by 10.}
\label{region2}
\end{figure}
\begin{figure}
\centering
\includegraphics[trim=0 -1cm 0 -0.5cm, clip, width=0.9\columnwidth]{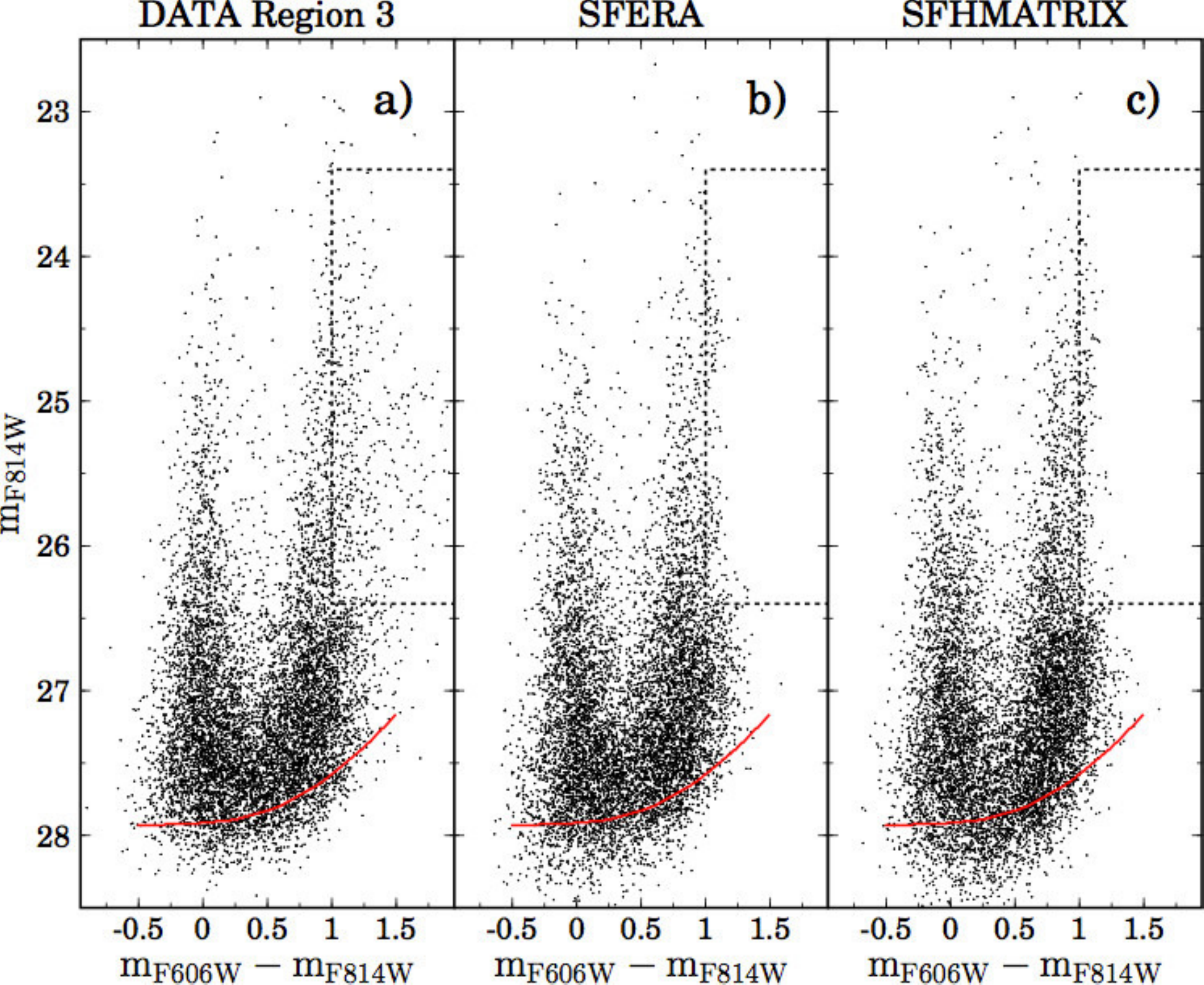}
\includegraphics[trim=0 0.4cm 0 1cm, clip, width=\columnwidth]{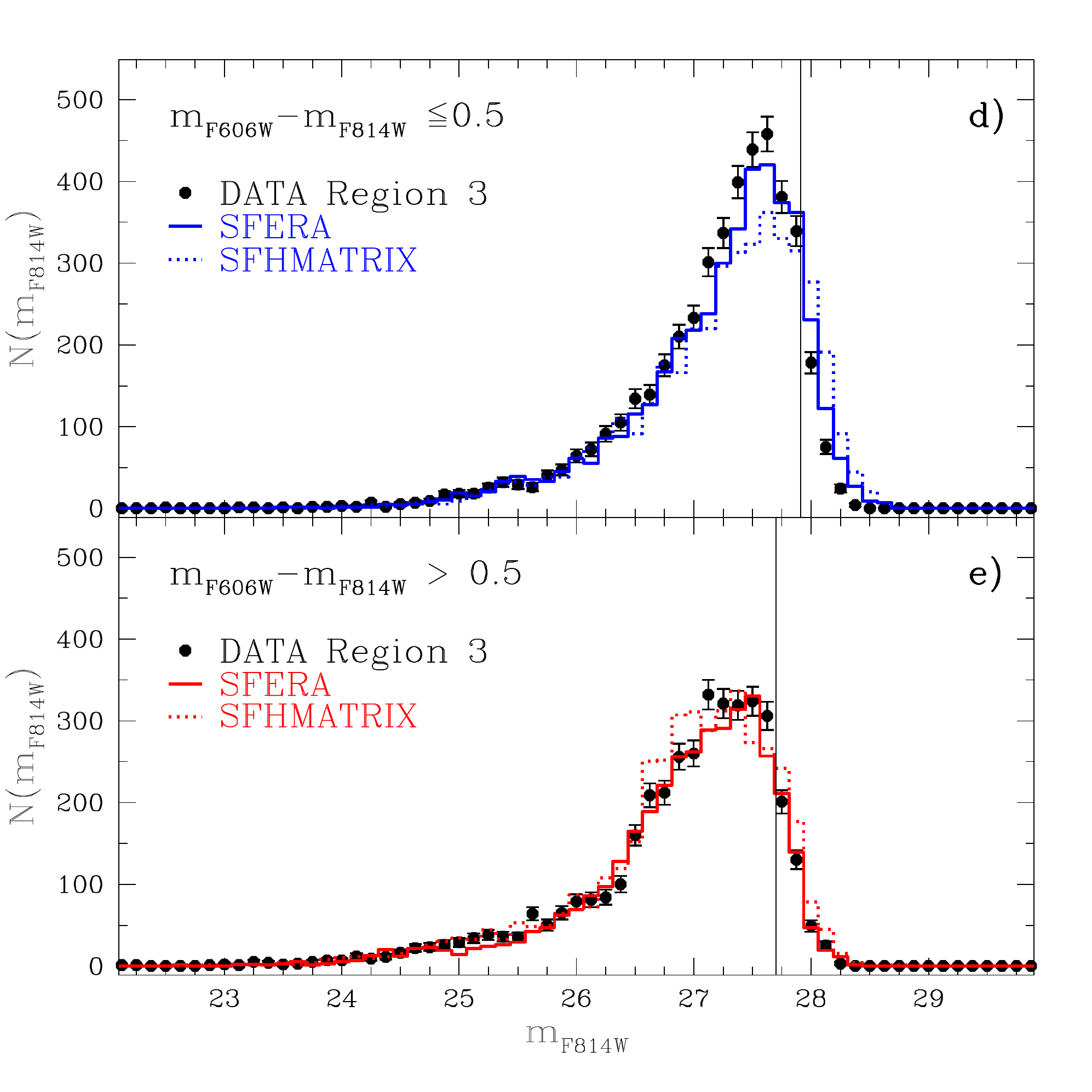}
\includegraphics[trim=0 0 0 0.5cm, clip, width=\columnwidth]{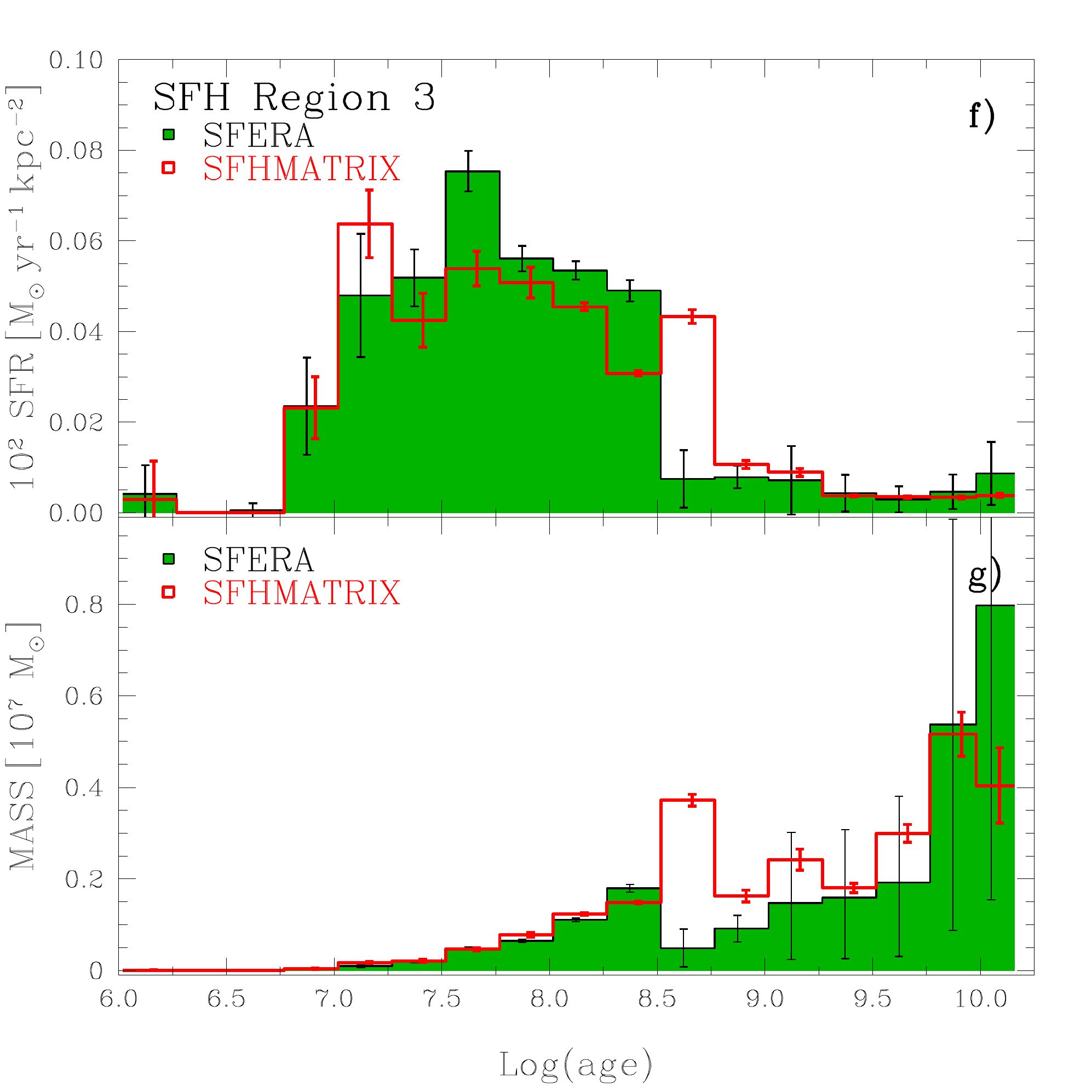}
\caption{Same as Figs. \ref{region1} and \ref{region2} but for Region 3. Notice that in this case the SFRs of panel f) are multiplied by 100.}
\label{region3}
\end{figure}
\begin{figure}
\centering
\includegraphics[trim=1.5cm 0 0 0, clip, width=0.93\columnwidth]{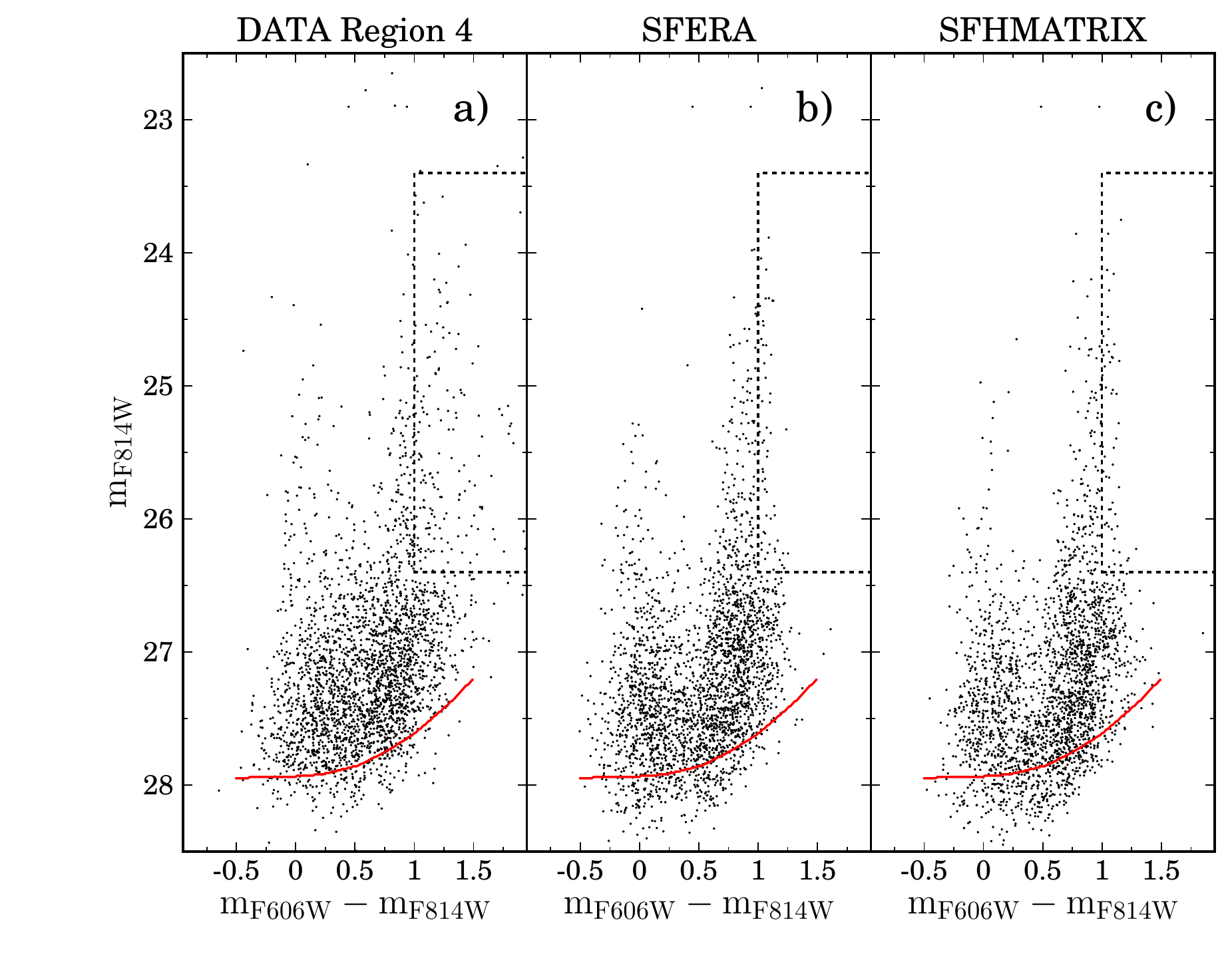}
\includegraphics[trim=0 0.4cm 0 1cm, clip, width=\columnwidth]{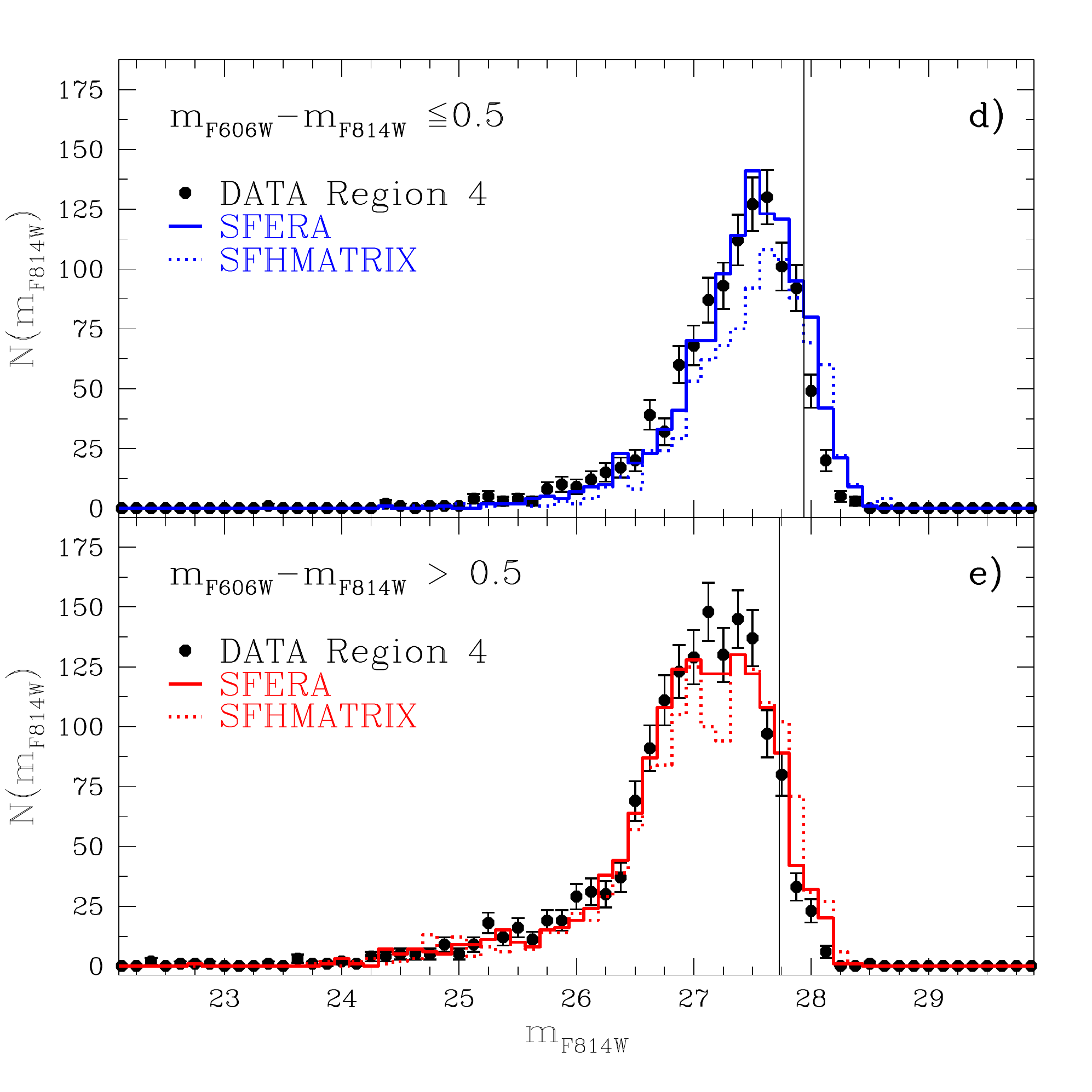}
\includegraphics[trim=0 0 0 0.5cm, clip, width=\columnwidth]{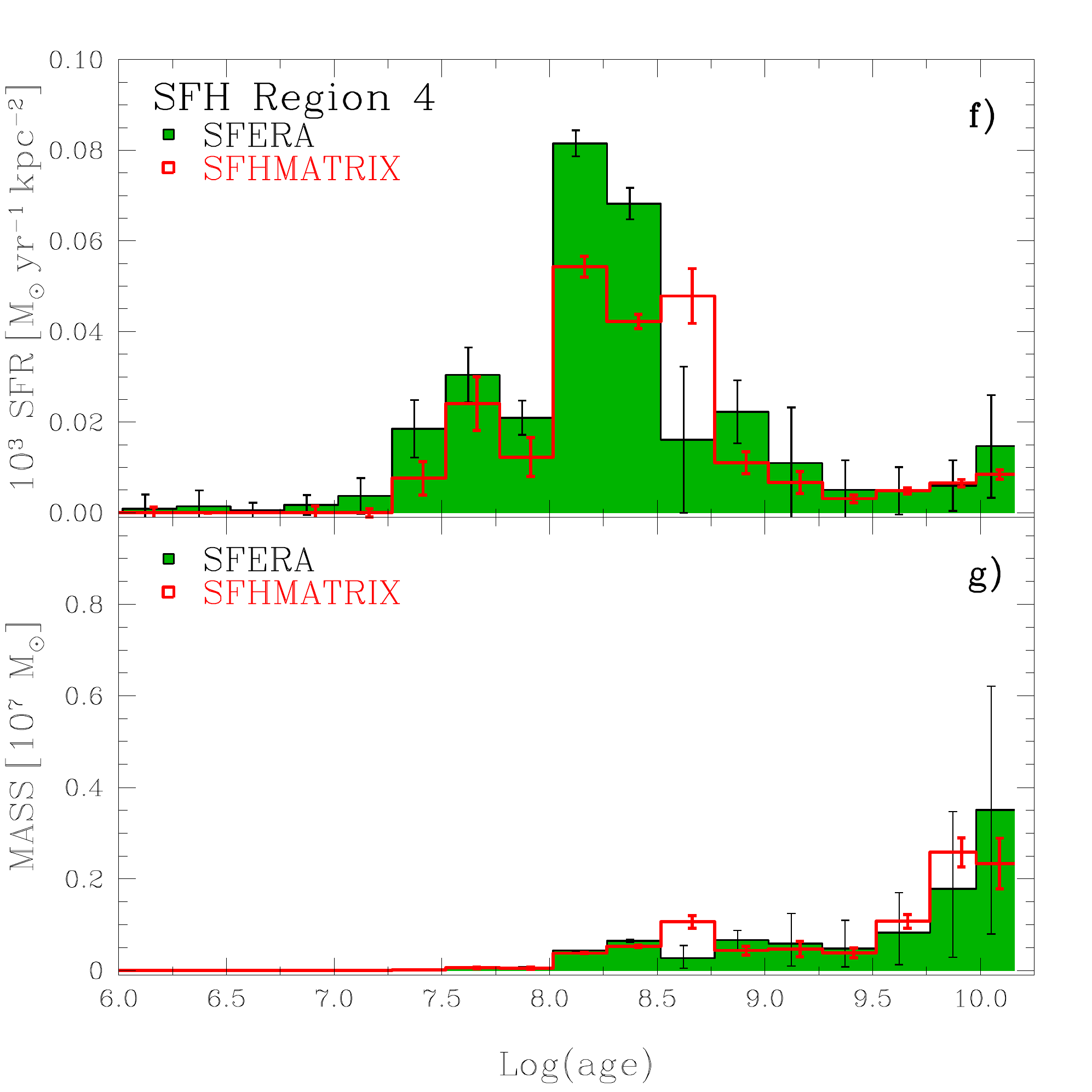}
\caption{Same as Figs. \ref{region1}$-$\ref{region3} but for Region 4. Notice that in this case the SFRs of panel f) are multiplied by 1000.}
\label{region4}
\end{figure}
Fig. \ref{region2} shows the results for Region~2. We excluded from the fit the region of the CMD where the completeness is below 50\%. Given the larger number of stars with respect to Region~1, we chose to use a pixel size of only 0.25 in both color and magnitude when running SFHMATRIX. We masked again the region of the TP-AGB phase. For SFERA we kept the same (variable) binning as for Region~1.

The blue LF in panel d) of Fig. \ref{region2} shows that SFHMATRIX under-predicts the counts at $27 < \mathrm{m_{F814W}} < 27.5$, while SFERA does so between 27 and 27.25, and both codes slightly over-predict the counts below 27.5; these regions correspond to the lower part of the MS where probably both the spatial resolution and the completeness of the photometry are degrading fast. For what concerns the red LFs ($\mathrm{m_{F606W}-m_{F814W}} > 0.5$), SFERA provides a quite good agreement in the brightest bins, and at $\sim \pm$ 0.5 magnitudes around the TRGB; on the other hand, we notice an under-prediction of counts at $25 < \mathrm{m_{F814W}} < 26$, where the (not-simulated) TP-AGB stars are located, and at $27 < \mathrm{m_{F814W}} < 27.5$. The SFHMATRIX code, instead, provides a better agreement in this magnitude range, but tends to over-predict the counts around the TRGB.

The SFH shows an overall trend very similar to that of Region~1, but the SFRs are almost 10 times lower (SFR$_{\, peak} \simeq 0.6 \times 10^{-2}$~M$_{\odot}$/yr/kpc$^2$ between 20 and 60~Myr) and were in fact multiplied  by a factor of $10$ in the plot to make them more visible. Also in this region we find the majority of the mass locked in the oldest ($> 1$~Gyr) stars, while there is no evidence for a significant recent activity.

This is the region where we find the largest differences between the results from the two codes, in particular in the formed stellar mass. 
In general, we consider SFERA more reliable, because it minimizes an actual likelihood using Poissonian statistics, and doesn't make he simplifying assumption of Gaussian statistics that is inherent to the $\chi^2$ minimization of SFHMATRIX. However, for the main characteristics of DDO~68 implied by our SFHs, the codes yield broadly consistent results (see Table \ref{tabella}).

\subsection{\textsc{Region 3}}
For Regions 3 and 4 of DDO~68 we kept the same binning of Region~2 and the same completeness limit of 50\%. As shown in Fig. \ref{region3}, this is the best reproduced region, both in the CMD and in the LFs. With SFERA there are no relevant discrepancies between the data and the models, while with SFHMATRIX there is an under-prediction of the blue counts at $\mathrm{m_{F814W}} \sim 27.5$ and an over-prediction of the red counts at $26.5 < \mathrm{m_{F814W}} < 27$.

The SFRs per unit area plotted in panel f) are multiplied  by a factor of $10^2$ to make them more visible; this is the region where a very recent SF activity is found by both codes. This is in agreement with the fact that the majority of the nebular emission is found in this region. The remaining evolution of the SF is quite similar to the other regions, but with lower rates (\mbox{SFR$_{\, peak} \simeq 7.5 \times 10^{-4}$~M$_{\odot}$/yr/kpc$^2$} between 20 and 60~Myr).

\subsection{\textsc{Region 4}} \label{sect_region4}
Fig. \ref{region4} shows the CMD, LFs and SFH of the most external region of the galaxy. There are very few stars here and a strong contamination from background galaxies. In fact, as expected, SFHMATRIX has some troubles in reproducing the right number of stars as can be noticed from the LFs. In both the blue and red LFs SFHMATRIX under-predicts the counts, also in the RGB region. SFERA instead reproduces well the blue LF, while it under-predicts the number of stars in the red LF at $27 < \mathrm{m_{F814W}} < 27.5$.
The plotted SFRs, also in this case, are multiplied by a factor of $10^3$, and in general are very low, showing a peak two bins older than the previous regions (SFR$_{\, peak} \simeq 0.8 \times 10^{-4}$~M$_{\odot}$/yr/kpc$^2$ between 100 and 200~Myr). This is not surprising since we expect this region to host mainly old populations of stars as discussed in Section \ref{pop_distr}. The oldest bins provide again the most relevant contribution to the formed stellar mass.

The total average SFR over the whole galaxy is $\simeq 0.01$~M$_{\odot}$/yr, and the corresponding astrated mass over the Hubble time is M$_{\star\,\mathrm{TOT}} \simeq 1.3 \times 10^8$ M$_{\odot}$.

Overall, the general good agreement between the solutions of SFHMATRIX and SFERA seems to suggest that the $\chi^2$ bias is not significantly affecting the results.
\begin{table*}
\caption{Summary of the results for SFRs and stellar mass in the four regions.}
\begin{center}
\begin{tabular}{cccccccc}
    \toprule
    \midrule
    \addlinespace[0.3em]
    \multirow{2}{*}{region} & \multirow{2}{*}{area [kpc$^2$]} & \multicolumn{2}{c}{$\langle$SFR$\rangle$ $_{\mathrm{age<1\,Gyr}}$ [M$_{\odot}$/yr/kpc$^2$]} & \multicolumn{2}{c}{M$_{\star\,\mathrm{TOT}}$ [$10^7$ M$_{\odot}$]} & \multicolumn{2}{c}{M$_{\star\ \mathrm{age>1\,Gyr}}$ [$10^7$ M$_{\odot}$]}\\
    \addlinespace[0.3em]
    & & SFERA & SFHMATRIX & SFERA & SFHMATRIX & SFERA & SFHMATRIX\\
    \midrule
    1 & $\ 1.1$ & $(7.91\pm 0.91)\times10^{-3}$ & $(9.68\pm 0.37)\times10^{-3}$ & $2.62\pm 1.04$ & $2.74\pm 0.20$ & $1.75\pm 1.03$ & $1.67\pm 0.19$\\
    2 & $\ 9.2$ & $(1.28\pm 0.12)\times10^{-3}$ & $(1.98\pm 0.03)\times10^{-3}$ & $6.94\pm 2.49$ & $7.85\pm 0.19$ & $5.76\pm 2.49$ & $6.03\pm 0.19$\\
    3 & $26.5$ & $(2.17\pm 0.20)\times10^{-4}$ & $(3.66\pm 0.07)\times10^{-4}$ & $2.41\pm 0.84$ & $2.61\pm 0.10$ & $1.83\pm 0.84$ & $1.64\pm 0.10$\\
    4 & $68.5$ & $(0.31\pm 0.05)\times10^{-4}$ & $(0.37\pm 0.02)\times10^{-4}$ & $0.93\pm 0.34$ & $0.94\pm 0.07$ & $0.72\pm 0.34$ & $0.68\pm 0.07$\\
    \midrule
    \bottomrule
\end{tabular}
\end{center}
\label{tabella}
\end{table*}

\subsection{\textsc{DDO 68 B}}
\begin{figure}
\centering
\includegraphics[width=\columnwidth]{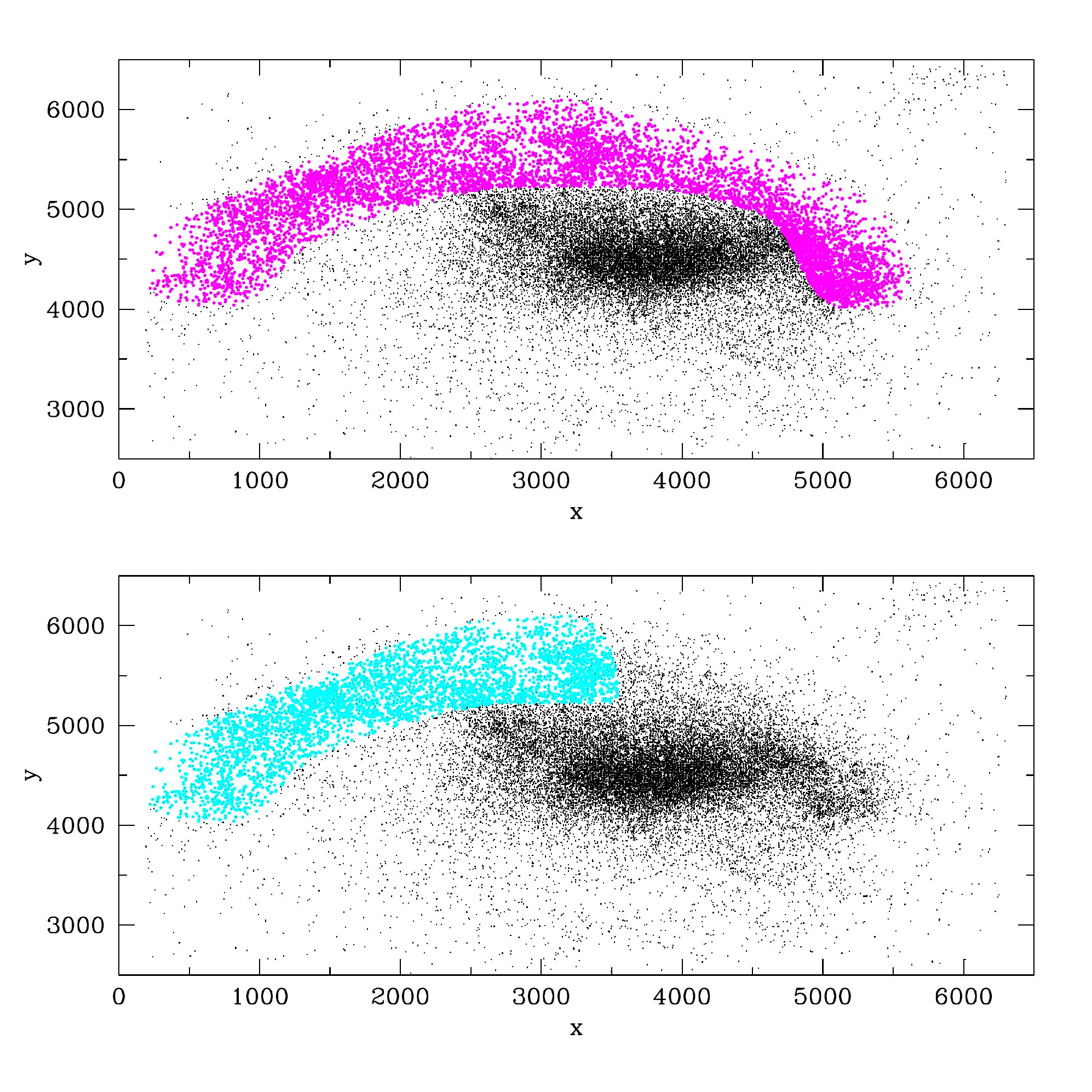}
\caption{Map of the stars in DDO~68 with enhanced in magenta and cyan two possible configurations of the arc-shaped structure identified as DDO~68~B and used for a separate derivation of the SFH.}
\label{arm_map}
\end{figure}
\begin{figure}
\centering
\includegraphics[trim=0 0 0 0.5cm, clip, width=\columnwidth]{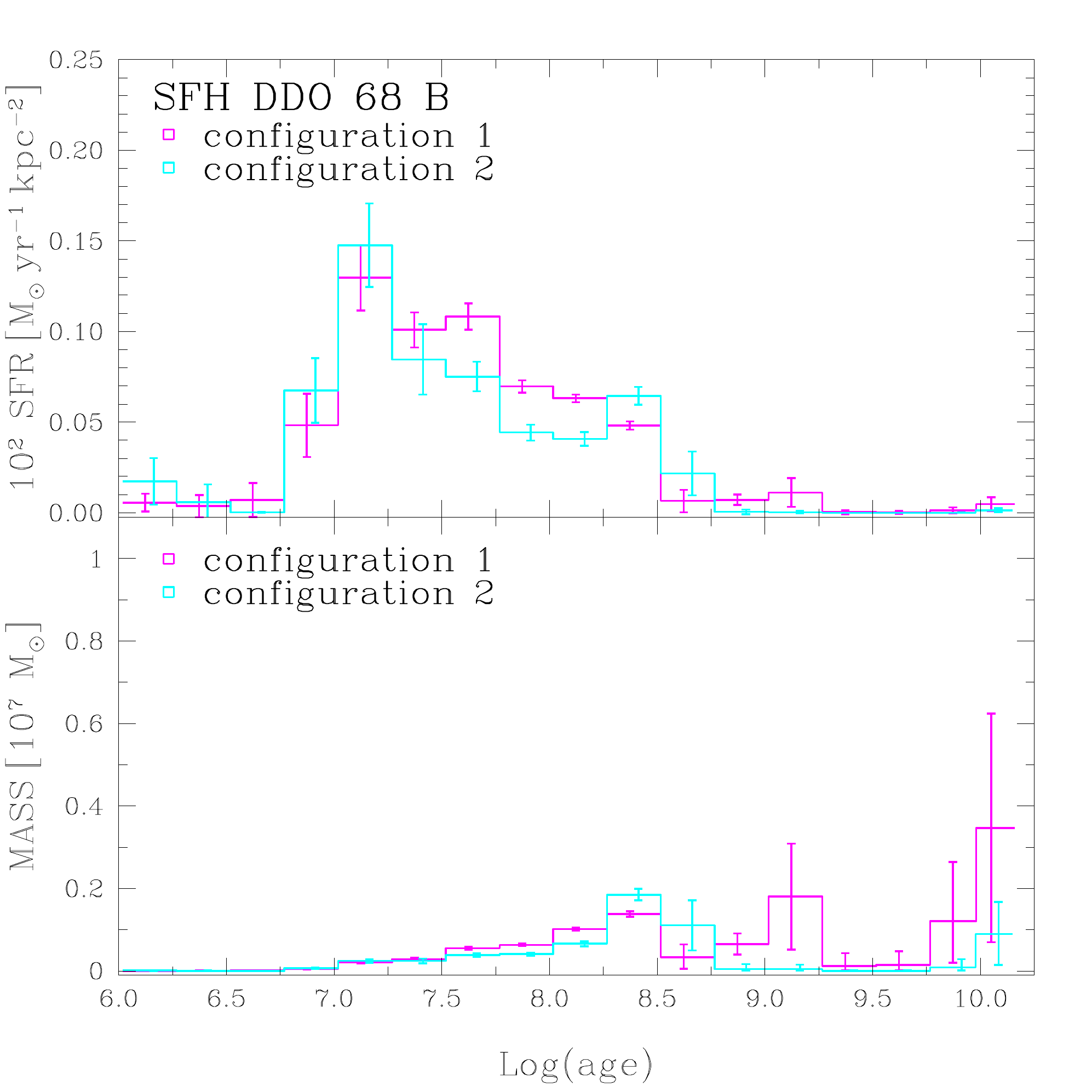}
\caption{SFHs for stars in magenta and cyan in Fig. \ref{arm_map}. Notice that in this case the SFRs are multiplied by 100.}
\label{arm_results}
\end{figure}
The peak in the star formation activity between $\sim 30$~Myr and $\sim 50$~Myr that we find in all regions of DDO~68 hints that this epoch may have been affected by a merging/accretion event between two different bodies, as suggested by \cite{Tikhonov2014}. For this reason we performed a separate analysis of the arc-shaped structure (dubbed by \citealt{Tikhonov2014} DDO~68~B) to explore also the possibility that the increased young star formation in Region 3 could be the result of this interaction.

In Fig. \ref{arm_map} we show two selections of stars falling in the Tail for which we re-derived the SFH. The results are shown in Fig. \ref{arm_results} (the SFRs per unit area are multiplied by a factor of $10^2$ as for Region 3). The SFH of the candidate accreted satellite in the last hundreds Myr is qualitatively similar to that of Region 2, that can be considered as representative of the main body, although the SF activity in the configuration 2 of the Tail is definitely more skewed towards more recent epochs, and close to absent at epochs older that 1 Gyr, as if the activity in the candidate satellite was mostly triggered when the interaction with the main body became strong. The young peak is the same as in Region 3, as expected since most of the H$\alpha$ emission is in common. However, the SFR normalized over the area in each bin is roughly double in the Tail with respect to the whole Region 3, since it has a smaller area but the majority of the young and intermediate stars. 

The average SFR over the last $\sim 1$~Gyr in the Tail Configuration 1 is $\langle$SFR$\rangle_{\mathrm{Tail\,1}}\simeq 3 \times 10^{-4}$~M$_{\odot}$/yr/kpc$^2$ while the total formed stellar mass is M$_{\star\,\mathrm{Tail\,1}} \simeq 1.2 \times 10^7$~M$_{\odot}$ that is $\sim 1/10$ of the stellar mass of the whole galaxy. In Configuration 2 they are $\langle$SFR$\rangle_{\mathrm{Tail\,2}}\simeq 4 \times 10^{-4}$~M$_{\odot}$/yr/kpc$^2$ and M$_{\star\,\mathrm{Tail\,2}} \simeq 0.6 \times 10^7$~M$_{\odot}$ respectively. The metallicity in the most recent bin of Configuration 2 is $Z\sim 0.0008$, so slightly higher than that of the other regions; however, the uncertainties on this determination prevent a firm conclusion.

The presence of recent SF in Region 3 is thus compatible with the merging/accretion scenario, that will be explored in more details in a forthcoming paper \citep{Annibali2016}.\\

\begin{figure*}
\includegraphics[trim=0 10cm 0 0, clip, width=\linewidth]{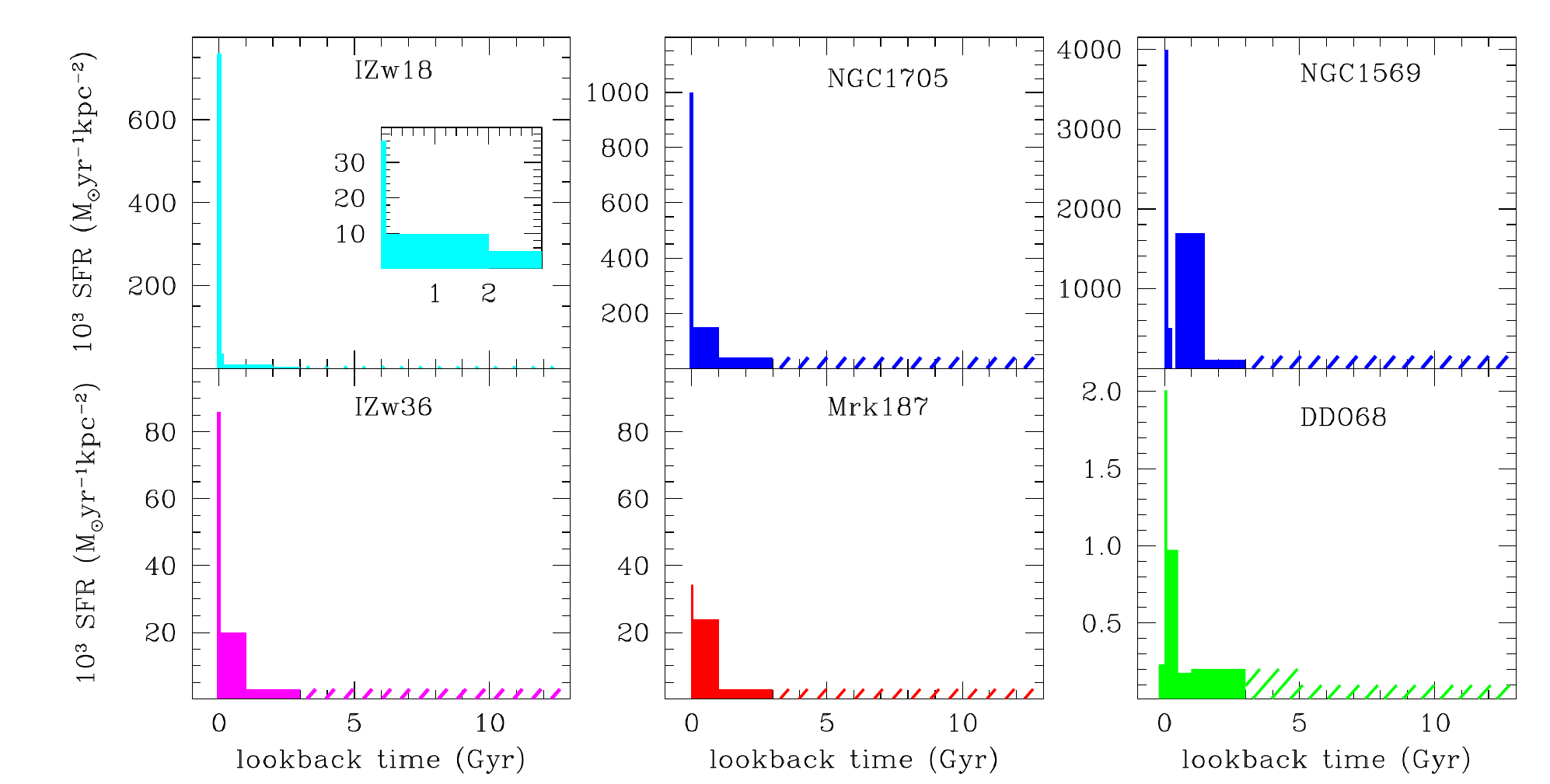}
\caption{SFHs of BCDs and starburst dwarfs in the Local Volume. The SFR densities are all multiplied by 1000, but the scale of the ordinate varies from one panel to the other. The onset in the top-left panel provides a blow-up of the plot portion between 0 and 3 Gyr ago, barely visible in the main graph. The solid histograms show what we consider a robust SFH resulting from the assumption that RGB stars resolved in these low metallicity galaxies are at least 3 Gyr old. The shaded areas are only indicative of  what is possibly the SFH at earlier epochs. Notice that for both NGC 1569 and NGC 1705 the plotted values result from the combination of two subsequent and consistent SFH derivations from the same group (see text for references).}
\label{sfall}
\end{figure*}
\section{Discussion and Conclusions}
In this paper, we presented several results on the star forming dwarf DDO~68, i.e. an accurate distance determination, a qualitative analysis of the spatial distribution of the different stellar populations, a detailed SFH based on the synthetic CMD method, and some very preliminary dynamical considerations. Here we summarize and discuss the main results obtained.

From the TRGB method, we derived a distance modulus of $(m-M)_0=30.41 \pm 0.12$~mag, i.e. a distance of $D=12.08 \pm 0.67$~Mpc. However, during our synthetic CMD analysis, we found more satisfactory using a slightly higher distance modulus, i.e. $(m-M)_0=30.51$, implying a distance of $D=12.65$~Mpc, still inside the errors of our TRGB determination, and closer to the distance of $D=12.74 \pm 0.27$~Mpc adopted by \cite{Cannon2014}.
From the presence of a well-populated RGB, we concluded that DDO~68 hosts a population of stars at least $1-2$~Gyr old (and possibly as old as a Hubble time), allowing us to reject the hypothesis that this is a young system experiencing its first burst of SF at recent epochs. The same conclusion was reached for the BCD I~Zw~18, with a similarly low metallicity, for which we clearly detected the RGB \citep{Aloisi2007}, and were able to provide a lower limit for the stellar mass formed at epochs older than $1-2$~Gyr ago \citep{Annibali2013}. Another galaxy worth mentioning in this context is Leo~P, a recently discovered star forming dwarf galaxy that has the same extremely low metallicity of DDO~68 and I~Zw~18. Leo~P is much closer (1.62 Mpc) and allows its oldest stars to be detected. In fact, \cite{McQuinn2015} find that it indisputably contains stars as old as 10~Gyr. As a matter of fact, there is no evidence so far for the existence of a young galaxy in the Local Universe.

From a quantitative derivation of the DDO~68's SFH based on the synthetic CMD method, we estimated that the mass locked up in old (age $>1$~Gyr) stars is M$_{\star} \simeq 1.0 \times 10^8$~M$_{\odot}$, i.e., almost 80\% of the total galaxy stellar mass. Obviously, the closer distance of DDO~68 compared to I~Zw~18 ($\sim 12$~Mpc versus $\sim 18$~Mpc) allowed us to better constrain the total stellar mass formed at these epochs. We consider this result quite robust, while the details on the SFH earlier than $1-2$~Gyr, as well as the age of the onset of the SF, should be taken with caution, due to the degeneracy of the stellar models on the RGB and to the large photometric errors at the faintest magnitudes. We emphasize however that the low metallicity of DDO~68 mitigates the age-metallicity degeneracy and allows us to argue on safer grounds that the redder RGB stars are likely $10-13$~Gyr old.
On the other hand, the SFH derived at epochs younger than $\sim 1$~Gyr offers a more realistic view of the recent SF activity in DDO~68. We derived an average SFR of $\simeq 7.1 \times 10^{-4}$~M$_{\odot}$~yr$^{-1}$~kpc$^{-2}$ over the last $\sim 1$~Gyr (we excluded Region 4 in this case because we want to estimate the average SFR over the main body, while Region 4 clearly includes the extremities and is the least well defined), which is typical of the majority of star-forming dwarf galaxies; the most recent SF activity is quite modest, and not even comparable to the burst strengths \citep[see e.g.][and references therein]{Tolstoy2009,Tosi2009} in ``monsters'' like NGC~1569 and NGC~1705.

Fig. \ref{sfall} shows the global SFR density (i.e. normalized to the area) of DDO~68 as a function of time, as derived for this paper, together with those of 5 other starburst dwarfs for which \textit{HST} allowed to derive the SFH within the look-back time given by the resolved RGB stars (formally $> 1$~Gyr, but in all cases taken as $\sim 3$~Gyr, thanks to the low metallicity of these galaxies, which significantly mitigates the age-metallicity degeneracy on the RGB). It is important to notice that at distances of few Mpc the SFH older than 1 Gyr necessarily relies on RGB stars, therefore the corresponding uncertainties are huge (see Appendix B of \citealt{Weisz2011}). Together with the mentioned NGC~1569 \citep{Greggio1998,Angeretti2005} and NGC~1705 \citep{Annibali2003,Annibali2009}, we include in this comparison I~Zw~18 \citep{Annibali2013}, I~Zw~36 \citep{Schulte-Ladbeck2001} and MrK~187 \citep{Schulte-Ladbeck2000}. The shown galaxies are too far away even to detect the He-burning phase of stars with ages between 0.5 and 1 Gyr, hence the resulting SFHs are rather uncertain also in that age range. However, within the uncertainties, the resulting scenario is correct. All these galaxies show a well defined very recent burst and moderate ancient SFRs, as old as the look-back time reached by the photometry, with no evidence for interruptions in the SF. Similar results are also found in other studies of dIrr galaxies \citep{McQuinn2010,Weisz2011} which typically show a significant SF in the oldest epochs plus, in the case of starburst galaxies, elevated levels of recent SF. The details of the SFHs, however, can considerably vary from galaxy to galaxy.

Extremely metal poor galaxies are recently being classified in two major groups, namely quiescent and active, in the attempt to understand their evolutionary stage and why they are so few \citep{James2015}. Quiescent ones (like Leo~P) are numerous but very faint, and therefore remain undetected when at relatively large distances, whereas active ones (like I~Zw~18) are undergoing a starburst and are easily detected even at large distances. DDO~68 is certainly not a starburst dwarf, but its level of star formation is high enough to let it be clearly recognizable in spite of its relatively large distance.

As with all studied star-forming dwarfs, DDO~68 does not show evidence for long interruptions in the SFH, with an overall activity more ``gasping'' than bursting; at all ages, the SFR density decreases from the center outwards. The distribution of the old stellar population is quite homogeneous and traces the whole galaxy, while young stars tend to be more concentrated. All these characteristics make DDO~68 quite a ``normal'' star forming dwarf galaxy.

Nonetheless, there are properties that make it exceptional: its highly disturbed morphology and the presence of an arc-shaped structure populated by stars of all ages, and rich in H~II regions, is a very peculiar characteristic which strongly suggests a merging/accretion scenario.  Interestingly, a  peak in the star formation activity between $\sim 30$~Myr and $\sim 50$~Myr is found in all regions of DDO~68, suggesting that this epoch may have been affected by the interaction of two merging bodies. A merging event between DDO~68 and a possible companion could also explain why DDO~68 is an outlier in the mass-metallicity relation for dwarf star-forming galaxies, with a metallicity similar to that of I~Zw~18, 
but with a stellar mass $\sim 10$ times higher: the H~II region abundances derived by \cite{Pustilnik2005} in the arc-shaped structure could in fact trace the metallicity of the accreted companion DDO~68~B, but not the (possibly higher) metallicity of DDO~68~A \citep{Tikhonov2014}. A merging event could also have triggered the recent SF observed at relatively large galactocentric distances (as demonstrated by our CMD analysis and by the presence of H$\alpha$ emission), which is a quite uncommon feature.
Finally, we recognized, in the RGB spatial map, a small concentration of stars toward the upper right edge of the ACS chip, noticed also by \cite{Tikhonov2014}; in order to better investigate the nature of this feature, we obtained new wide-field imaging data at the Large Binocular Telescope (LBT). These new data, together with preliminary results on the accretion history of DDO~68, will be presented in a forthcoming paper \citep{Annibali2016}.

DDO~68 is also part of the LEGUS (Legacy Extra Galactic UV Survey) program, which is an HST Treasury Program imaging 50 local (within $\sim 12$~Mpc) galaxies with WFC3 and ACS with the goal of studying in detail the most recent SF, the effects of the SFH on the UV SFR calibrations, the impact of environment on the star formation, and the properties and the evolution of star clusters. Thanks to these new data, we will be able to explore the very recent SF in DDO~68 and to better understand what triggered its most recent SF activity.

\acknowledgments{
These data are associated to the HST GO Program 11578 (PI A. Aloisi). Support to this program was provided by NASA through a grant from the Space Telescope Science Institute. E.S. is supported by INAF through a Ph.D. grant at the University of Bologna. F.A. was supported through PRIN-MIUR-2010 LY5N2T that partially funded also E.S. and M.T.
E.S. is grateful to the Space Telescope Science Institute (STScI) for the kind hospitality and the tools provided during the period spent in Baltimore.
}

\bibliography{bib}

\end{document}